\documentclass[a4paper,11pt]{article}
\pdfoutput=1 

\usepackage{jheppub} 
\usepackage{physics,slashed,xcolor}
\usepackage{cleveref}
\usepackage{hhline}
\usepackage{hyperref}
\usepackage[T1]{fontenc} 
\usepackage[normalem]{ulem}
\usepackage{mathtools}

\def\cevns{CE$\nu$NS~}

\newcommand{\beq}{\begin{equation}}
\newcommand{\eeq}{\end{equation}}

\newcommand{\ber}{\begin{eqnarray}}
\newcommand{\eer}{\end{eqnarray}}

\title{\boldmath Flavor-dependent radiative corrections in coherent elastic neutrino-nucleus scattering}

\author[a,b,1]{Oleksandr Tomalak,\note{Corresponding author.}}
\author[b]{Pedro Machado,}
\author[c,2]{Vishvas Pandey,\note{Present Address: Fermi National Accelerator Laboratory, Batavia, Illinois 60510, USA}}
\author[a,b]{Ryan Plestid}

\affiliation[a]{Department of Physics and Astronomy, University of Kentucky, Lexington, KY 40506, USA}
\affiliation[b]{Theoretical Physics Department, Fermilab, Batavia, IL 60510, USA}
\affiliation[c]{Department of Physics, University of Florida, Gainesville, FL 32611, USA}
\emailAdd{oleksandr.tomalak@uky.edu}
\emailAdd{pmachado@fnal.gov}
\emailAdd{vpandey@fnal.gov}
\emailAdd{rpl225@uky.edu}

\preprint{FERMILAB-PUB-20-524-T}
\abstract{
We calculate coherent elastic neutrino-nucleus scattering cross sections on spin-0 nuclei (e.g.\ $^{40}$Ar and $^{28}$Si) at energies below 100 MeV within the Standard Model and account for all effects of permille size. We provide a complete error budget including uncertainties at nuclear, nucleon, hadronic, and quark levels separately as well as perturbative error. Our calculation starts from the four-fermion effective field theory to explicitly separate heavy-particle mediated corrections (which are absorbed by Wilson coefficients) from light-particle contributions. Electrons and muons running in loops introduce a nontrivial dependence on the momentum transfer due to their relatively light masses. These same loops, and those mediated by tau leptons, break the flavor universality because of mass-dependent electromagnetic radiative corrections. Nuclear physics uncertainties significantly cancel in flavor asymmetries resulting in subpercent relative errors. We find that for low neutrino energies, the cross section can be predicted with a relative precision that is competitive with neutrino-electron scattering. We highlight potentially useful applications of such a precise cross section prediction ranging from precision tests of the Standard Model, to searches for new physics and to the monitoring of nuclear reactors.}

\makeatletter
\gdef\@fpheader{}
\makeatother

\begin{document} 
\maketitle
\flushbottom

\section{Introduction \label{intro}}

Coherent elastic neutrino-nucleus scattering~\cite{Stodolsky:1966zz,Freedman:1973yd,Kopeliovich:1974mv}, or \cevns for short, is the most recently discovered form of neutrino interaction with matter~\cite{Akimov:2017ade}.
In these interactions, the momentum transferred to a system is sufficiently low, such that the neutrino probes the nucleus as a whole, instead of distinguishing individual nucleons.
Compared to the usual neutrino-nucleon interactions, the most important feature of \cevns is what has been dubbed the ``coherent enhancement'': the cross section is proportional to the square of the weak charge of the nucleus.
This enhances the \cevns interaction rate 10- to 100-fold, depending on the nucleus in question relative to incoherent cross sections.

This new way of detecting neutrinos has attracted a great deal of attention from the high-energy community. 
The possibility of doing neutrino physics with relatively small detectors in the kg~\cite{violeta,neutrino-2020-poster-1,neutrino-2020-poster-2,neutrino-2020-poster-3,Flores:2021ihw} to tonne~\cite{Tayloe:2017edz,Akimov:2020xsa,Kumpan:2020snb,CCM} scale, as opposed to 100-tonne~\cite{Antonello:2015lea,Machado:2019oxb,Abratenko:2019jqo,McConkey:2017dsv} or even kiloton detectors~\cite{Farnese:2019xgw,Antonello:2015lea,Machado:2019oxb}, opens up the possibility of competitively measuring neutrino properties with small-scale projects~\cite{Aguilar-Arevalo:2019jlr,neutrino-2020-poster-1,neutrino-2020-poster-3,Fernandez-Moroni:2020yyl,Galindo-Uribarri:2020huw}. \cevns provides an unavoidable background floor for dark matter direct detection~\cite{Cabrera:1984rr,Drukier:1986tm,Monroe:2007xp,Vergados:2008jp,Strigari:2009bq,Gutlein:2010tq,Harnik:2012ni,Billard:2013qya,OHare:2016pjy,Cerdeno:2016sfi,Bertuzzo:2017tuf,Boehm:2018sux,Gelmini:2018ogy,Papoulias:2018uzy}. Moreover, as \cevns is inherently a low-energy process, it provides a natural window to study light, weakly-coupled, new physics in the neutrino sector~\cite{Amanik:2004vm,Barranco:2005yy,Scholberg:2005qs,Barranco:2007tz,Barranco:2008rc,Barranco:2009px,Formaggio:2011jt,Espinoza:2012jr,Billard:2014yka,Dutta:2015nlo,Lindner:2016wff,Akimov:2017ade,Coloma:2017ncl,Kosmas:2017tsq,Liao:2017uzy,AristizabalSierra:2017joc,Farzan:2017xzy,Canas:2017umu,Boehm:2018sux,Billard:2018jnl,Altmannshofer:2018xyo,AristizabalSierra:2018eqm,Brdar:2018qqj,Abdullah:2018ykz,Farzan:2018gtr,Denton:2018xmq,Gonzalez-Garcia:2018dep,Esteban:2018ppq,AristizabalSierra:2019zmy,Miranda:2019skf,Alikhanov:2019drg,Bell:2019egg,Bischer:2019ttk,AristizabalSierra:2019ufd,Dev:2019anc,Khan:2019cvi,Cadeddu:2019eta,Giunti:2019xpr,Akimov:2019rhz,Han:2019zkz,Aguilar-Arevalo:2019zme,Baxter:2019mcx,Papoulias:2019xaw,Coloma:2019mbs,Babu:2019mfe,Canas:2019fjw,Dent:2019ueq,Dutta:2019eml,Miranda:2019wdy,Esteban:2020opq,Denton:2020hop,Dutta:2020che,Chang:2020jwl,Flores:2020lji,Akimov:2020pdx,Abdullah:2020iiv,Li:2020lba,Amaral:2020tga,Abdallah:2020biq,Sadhukhan:2020etu,Suliga:2020jfa,Sinev:2020bux,Skiba:2020msb,Dutta:2020enk,Denton:2020uda,Ding:2020uxu,Cadeddu:2020nbr,Miranda:2020syh,Snowmass2021LoICoherent,Aprile:2020thb,Wei:2020rbm}. Nevertheless, taking advantage of \cevns to do precision neutrino physics is not without its challenges.

Besides obvious experimental challenges, such as detecting low nuclear recoils, suppressing backgrounds sufficiently or reconstructing the incoming neutrino energy, capitalizing on \cevns requires overcoming certain theoretical difficulties.
In this manuscript, we deal with a few of these.
We present a calculation of radiative corrections that is universal to all neutral-current processes that may affect the \cevns cross section at the few \% level. This is required to properly interpret future precision physics studies with CE$\nu$NS, including weak mixing angle~\cite{Scholberg:2005qs,Canas:2018rng,Khan:2019cvi,Huang:2019ene,Miranda:2019skf,Cadeddu:2019qmv,Cadeddu:2019eta,Baxter:2019mcx,Papoulias:2019xaw,Fernandez-Moroni:2020yyl,Wei:2020rbm} and neutrino charge radius~\cite{Bernabeu:2002nw,Bernabeu:2002pd,Papavassiliou:2005cs,Kosmas:2017tsq,Cadeddu:2018dux,Cadeddu:2019eta,Khan:2019cvi,Baxter:2019mcx,Papoulias:2019xaw} extractions. 
We also calculate the flavor-dependent corrections to this process, which can change the \cevns cross section at the few \% level. These flavor-dependent contributions are particularly interesting because they allow a neutral-current process to statistically distinguish the neutrino flavor. While small, the flavor dependence of neutral-current scattering is \emph{thresholdless} (being dictated by experimental limits) and permits the detection of neutrino flavor independent of the associated charged lepton's mass. Remarkably, this flavor sensitivity implies that the tau neutrino can be accessed at neutrino energies of order $~10~\mathrm{MeV}$, well below the tau production threshold in neutrino-nucleon scattering of $E_\nu \sim 3.5~\mathrm{GeV}$. \cevns offers a realistic avenue with which to observe these effects since its cross section is much larger than other neutral-current processes such as neutrino-electron scattering. 

As mentioned above, despite \cevns being a neutral-current process, it receives electromagnetic radiative corrections which are naively of $\mathrm{O}(\alpha/\pi \sim 0.2\%)$, however they can be enhanced by kinematic factors resulting in percent-level corrections to tree-level results. The era of precision \cevns detectors, from the perspective of the Standard Model, is therefore defined by $\mathrm{O}(1\%)$ precision commensurate with the optimistic projections for next-generation detectors mentioned above, see e.g.\ the future physics goals outlined in~\cite{Akimov:2018ghi}.

Leading-order radiative corrections to neutrino neutral-current processes differ qualitatively from charged-current processes or parity-violating electron-nucleus scattering in that no $W-\gamma$ or $Z-\gamma$ box diagrams appear (because neutrinos are neutral). This means that at $\mathrm{O}(\mathrm{G}_\mathrm{F}^2 \alpha)$ radiative corrections simply induce a photon mediated interaction between the neutrino and the nucleus such that nuclear physics enters only through on-shell matrix elements of hadronic currents.
In particular, radiative corrections to \cevns are proportional to $\mel{A(p')}{\hat{J}_\text{EM}}{A(p)}$, where $|A(p)\rangle$ denotes the nucleus state and $\hat J_\text{EM}$ is the electromagnetic current operator, which is directly accessible via high-precision elastic electron scattering data. Consequently, all dominant nuclear physics uncertainties are contained in the weak form factor $\mathrm{F}_\mathrm{W}$ which is less well known~\cite{Amanik:2007ce,Amanik:2009zz,Huang:2019ene,AristizabalSierra:2019zmy,Khan:2019cvi,Canas:2019fjw,Co:2020gwl,Coloma:2020nhf,VanDessel:2020epd,Hoferichter:2020osn,Cadeddu:2019eta}.

\begin{figure}[tbp]
\centering 
\includegraphics[height=0.386\textwidth]{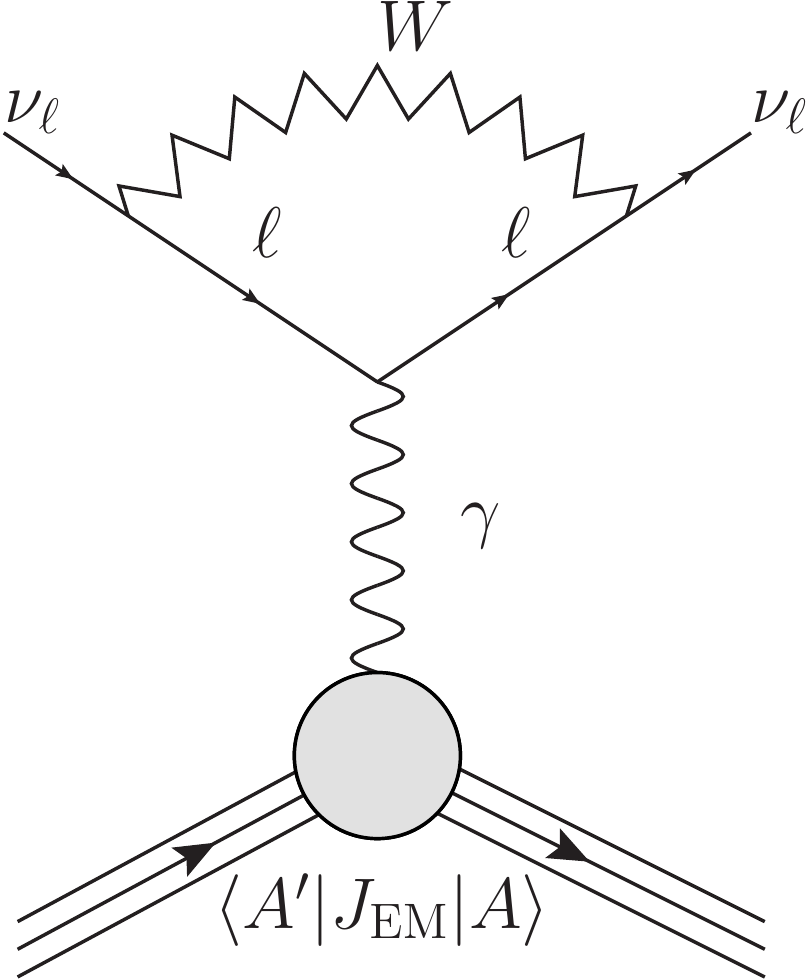}
\hspace{65pt}
\includegraphics[height=0.35\textwidth]{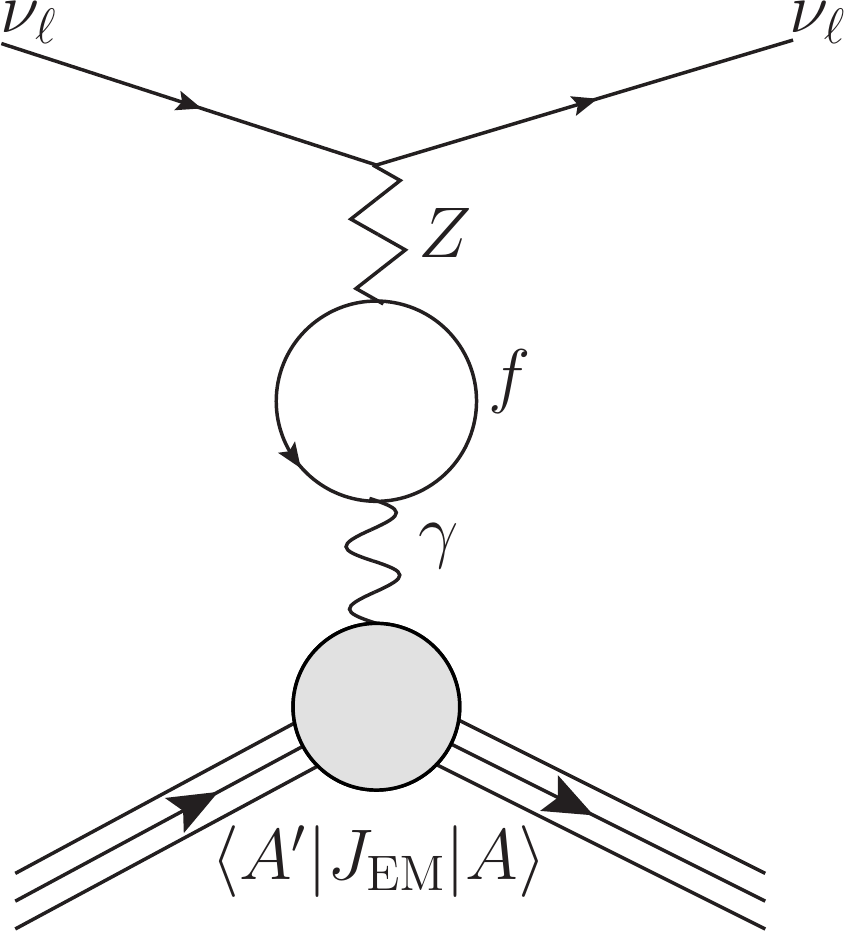}
\caption{\label{WZ-diags} Radiative corrections induce a photon-mediated interaction with nuclei. Charged currents (left diagram) lead to flavor-dependent calculable corrections. Neutral currents (right diagram) lead to flavor-independent corrections, some of which involve light quarks and are inherently nonperturbative. Rather than working directly with the Standard Model Lagrangian, we use a more efficient four-fermion treatment~\cite{Tomalak:2019ibg,Tomalak:2019fls,Hill:2019xqk} presented in~\cref{4-fermi}, that incorporates all electroweak physics and effects due to running coupling constants and heavy-particle loops inside Wilson coefficients. }
\end{figure}

\newpage
In contrast to corrections from nuclear effects (e.g.\ in the weak form factor), radiative corrections introduce qualitatively new ingredients to \cevns that are not present at leading order. For example, \cevns is often idealized as a flavor-independent process, however this is only true at tree level. Loops with charged leptons depend on the mass of the lepton resulting in flavor-dependent corrections (see~\cref{WZ-diags}). These effects are reasonably well appreciated in the \cevns literature at a qualitative level~\cite{Sakakibara:1979rc,Botella:1986wy,Cadeddu:2018dux,Kosmas:2015vsa,Vilain:1994hm,Scholz:2017ldm}. According to~\cite{Kayser:1982br,Shrock:1982sc}, neutrino-photon interactions are induced via loops of charged particles. As for any spin-$1/2$ particle, the on-shell vertex can be parameterized using Dirac spinors and form factors $\bar{u}(p_2)\Gamma_\mu(Q^2) u(p_1)$ with $Q^2 = - \left( p_1 - p_2 \right)^2$ (see e.g.~\cite{Giunti:2014ixa} for a review in the context of neutrinos). Extracting the form factor slope from the experimental data, one can probe a conventionally defined neutrino charge radius~\cite{Bernabeu:2002pd}. While we agree that \cevns is a probe of the neutrino's electromagnetic properties, it probes the $Q^2$ dependence of form factors rather than their $Q^2\rightarrow 0$ limit. 

For concreteness, most of our discussion focuses on neutrinos from a pion decay-at-rest source. 
We comment on other sources of low-energy neutrinos when needed. 
Although typical momentum transfers $Q^2\sim E_\nu^2\sim (30~\text{MeV})^2$ are small relative to the scales relevant for nuclear coherence, they are large relative to the scales controlling quantum fluctuations, namely $m_e$ and $m_\mu$. In fact, the $Q^2\rightarrow 0$ limit can only be taken safely for virtual $\tau$ loops and for loops with $\mu$ over some kinematic range, as we will see later.
More precisely, the flavor-dependent contribution is proportional to the vacuum polarization function in QED, and depends on both $Q^2/m_e^2$ and $Q^2/m_\mu^2$ neither of which is small for e.g.\ a neutrino energy of $E_\nu=50$ MeV. 
In the \cevns literature, it is often claimed that the effect of lepton loops can be included in cross section calculations via a prescriptive replacement of the Weinberg angle by an effective value. 
This prescription assumes a strict $Q^2\rightarrow 0$ limit, however (as outlined above) this condition is not always satisfied in \cevns as well as in neutrino-electron scattering at kinematics of modern accelerator-based neutrino experiments~\cite{Tomalak:2019ibg,Tomalak:2019fls}. We provide a general treatment for $Q^2$ dependence of radiative corrections in this paper.

In what follows, we study radiative corrections to \cevns working to $\mathrm{O}(\mathrm{G}_\mathrm{F}^2\alpha)$, i.e. next-to-leading order (NLO). We emphasize that flavor differences $ \sigma_{\nu_\ell} - \sigma_{\nu_{\ell'}}$ are calculable at $\mathrm{O}(\mathrm{G}_\mathrm{F}^2\alpha^2)$ with the same treatment of nuclear uncertainties as the leading-order \cevns cross section. For NLO corrections (both flavor-dependent and independent), we provide a full error budget for radiative corrections, while also accounting for nuclear- and nucleon-level uncertainties at the same requisite $\mathrm{O}(1\%)$ level of precision. 

The rest of this paper is organized as follows: In~\cref{cevns-NLO}, we introduce the relevant theory for \cevns scattering on spin-0 nuclei. In~\cref{4-fermi}, we begin with a discussion of \cevns kinematics and a discussion of nuclear form factors. We then describe the four-Fermi effective field theory (EFT), as outlined in~\cite{Hill:2019xqk}, that is used in the rest of the paper. In~\cref{nucleons,nucleus}, we make use of this EFT framework to give self-consistent definitions of the nucleon and nuclear form factors including the weak nuclear charge. In~\cref{error-budget}, we provide a comprehensive error budget for the cross section. In~\cref{flavor-dep}, we focus on the flavor dependence and specifically the flavor difference of cross sections, the flavor asymmetry, defined as $\left( \sigma_{\nu_\ell}-\sigma_{\nu_{\ell'}} \right)/\sigma_{\nu_{\ell}}$. In~\cref{NLO-diff}, we discuss how the flavor asymmetry can be computed at NLO, i.e. $\mathrm{O}(\mathrm{G}_\mathrm{F}^2\alpha^2)$, with substantially reduced nuclear uncertainties (see~\cref{error-buget-asym}). We then briefly sketch possible useful applications of our results and discuss the future of \cevns in~\cref{applications}. Finally, in~\cref{conclusions}, we summarize our findings and reiterate the applicability of our work to the future \cevns program. 

\section{Coherent elastic neutrino-nucleus scattering (CE$\nu$NS) on spin-0 nuclei \label{cevns-NLO}}
We focus here on spin-0 nuclei both because they are simple (our focus is on radiative corrections rather than nuclear physics) and because at least two nuclear targets of spin-0 are relevant to upcoming \cevns detectors. 
Liquid argon is used as a common liquid noble detector material all of whose stable isotopes ($^{40}$Ar, $^{38}$Ar, and $^{36}$Ar) are spin-0, and silicon is the main material in Skipper-CCDs~\cite{Moroni:2011xs, Tiffenberg:2017aac,Aguilar-Arevalo:2019jlr, violeta}, two of whose stable isotopes ($^{28}$Si and $^{30}$Si) are spin-0 and compose 95\% of silicon's natural abundance. 
Moreover, because a detailed understanding of $^{40}$Ar's nuclear physics is essential for the DUNE physics program~\cite{Alion:2016uaj,Abi:2020evt}, we are optimistic that the theoretical nuclear uncertainties relevant for \cevns will be steadily improved in coming years.

At tree level, neutrino neutral-current scattering on spin-0 nuclei can be described by a single form factor. At next-to-leading order (NLO) in the electromagnetic coupling constant $\alpha$, photon-mediated scattering takes place and the cross section inherits a flavor-dependent contribution entering with a charge form factor of the nucleus\footnote{Note that the tree-level differential cross section can be obtained setting radiative correction to zero. Neglecting nuclear and nucleon structure dependence, it is simply 
\begin{align}
    \frac{\mathrm{d} \sigma_{\nu_\ell}}{\mathrm{d} T}
    &\longrightarrow \frac{\mathrm{G}_\mathrm{F}^2 M_\mathrm{A}}{4\pi} \left( 1 -\frac{M_\mathrm{A} T}{2E_\nu^2} \right) \left[N-(1-4\sin^2\theta_\mathrm{W})Z\right]^2,
\end{align}
where $N$ and $Z$ are the number of neutrons and protons in the nucleus and $\theta_\mathrm{W}$ is the Weinberg angle.}
\begin{equation}\label{cevns-cross-section}
    \frac{\mathrm{d} \sigma_{\nu_\ell}}{\mathrm{d} T} = \frac{\mathrm{G}_\mathrm{F}^2 M_\mathrm{A}}{4\pi} \left( 1-\frac{T}{E_\nu} -\frac{M_\mathrm{A} T}{2E_\nu^2} \right) \left( \mathrm{F}_\mathrm{W} \left(Q^2\right)  + \frac{\alpha}{\pi} \qty[\delta^{\nu_\ell} + \delta^\text{QCD}]   \mathrm{F}_\mathrm{ch}(Q^2)\right)^2,
\end{equation}
with incoming neutrino energy, $E_\nu$, recoil nucleus kinetic energy, $0 \le T \le 2 E_\nu^2/(M_\mathrm{A} + 2 E_\nu)$, and the mass of the nucleus, $M_\mathrm{A}$. The expression depends on the weak, $\mathrm{F}_\mathrm{W}$, and charge, $\mathrm{F}_\mathrm{ch}$, nuclear form factors. The charge form factor enters multiplied by $\delta^{\nu_\ell}$ and $\delta^\text{QCD}$ which are radiative corrections defined below in~\cref{eq:radiative_correction-2,eq:radiative_correction-3}. The four-momentum transfer $Q^2$ can be conveniently expressed as $Q^2 = 2 M_\mathrm{A} T$. The corrections induced by hadronic and/or quark loops, proportional to $\delta^\text{QCD}$, are flavor independent, whereas the corrections from charged leptons, proportional to $\delta^{\nu_\ell}$, depend on the neutrino flavor $\ell$.

\cevns is a low-energy process. To describe the cross sections at a percent level of precision or better, it is important to properly account for the running of Lagrangian parameters from the weak scale down to the scales relevant for CE$\nu$NS. In the Standard Model (SM), this is a cumbersome task, and it is much more efficient to work within the four-fermion EFT when $W,~Z$, and $h$ are explicitly integrated out. This drastically reduces the number of contributing Feynman diagrams, while allowing for a full and systematic treatment of any loop-induced corrections from heavy particles.

Such an EFT approach has been worked out in detail by one of us in~\cite{Hill:2019xqk}, and has been successfully applied to neutrino-electron scattering in~\cite{Tomalak:2019ibg}. This latter process and \cevns share many similar features. In what follows, we introduce the EFT description appropriate to low-energy neutral-current interactions (such as \cevns or elastic $\nu e$ scattering). We summarize how heavy-quark and lepton contributions can be included perturbatively, whereas light-quark contributions require a nonperturbative treatment at CE$\nu$NS kinematics. The nonperturbative corrections are flavor independent, such that \cevns flavor differences are not affected by corresponding errors.

\subsection{Four-fermion effective field theory and radiative corrections\label{4-fermi}}

Neutrino-photon interactions scale with the photon momentum as $Q^2$ and can be captured by a dimension-six operator $\partial_\lambda F^{\lambda \rho} \bar{\nu}_\ell \gamma_\rho \mathrm{P}_\mathrm{L} \nu_\ell$. Following~\cite{Hill:2019xqk}, this interaction can be conveniently removed via a field redefinition (see~\cref{field-redefinition} for a detailed discussion), leading to 
\begin{align}  \label{effective_Lagrangian_quarks}
    {\cal L}_{\rm eff} &\supset - \sum_{\ell,\ell^\prime}  \bar{\nu}_\ell \gamma^\mu \mathrm{P}_\mathrm{L} \nu_\ell
    \, \bar{\ell}^\prime \gamma_\mu (c_\mathrm{L}^{\nu_\ell \ell^\prime} \mathrm{P}_\mathrm{L}
    + c_\mathrm{R}^{\nu_\ell \ell^\prime} \mathrm{P}_\mathrm{R}) \ell^\prime
    - \sum_{\ell,q}  \bar{\nu}_\ell \gamma^\mu \mathrm{P}_\mathrm{L} \nu_\ell \,
    \bar{q} \gamma_\mu (c_\mathrm{L}^{q} \mathrm{P}_\mathrm{L} + c_\mathrm{R}^{q} \mathrm{P}_\mathrm{R}) q 
    \nonumber\\
    & \quad\quad -\frac14 F_{\mu\nu}F^{\mu\nu} + e \sum_{\ell} Q_\ell \bar{\ell} \gamma_\mu \ell A^\mu + e \sum_q Q_q \bar{q}\gamma_\mu q A^\mu ~,
\end{align}
with projection operators on left- and right-handed chiral states $\mathrm{P}_{\mathrm{L}, \mathrm{R}} = \frac{1\mp \gamma_5}{2}$, and electric charges $Q_\ell,~Q_q$ being taken in units of the proton charge. The Wilson coefficients appearing here are evaluated in the $\overline{\text{MS}}$ renormalization scheme at a scale $\mu \gtrsim 2$ GeV appropriate for a quark-level description.
In this Lagrangian, the photon-neutrino couplings have been removed explicitly by the field redefinition discussed above, and shuffled into the contact interaction Wilson coefficients. By doing this, any diagram involving photon exchange with the nucleus that is mediated by heavy degrees of freedom (e.g.\ the top or bottom quark) is shuffled into the left- and right-handed quark couplings. This explicitly decouples low- and high-energy degrees of freedom, and results in a modified definition of the weak nuclear form factor (see~\cref{eq:form_factor_weak}). 
Such effects are proportional to the nuclear charge form factor $\mathrm{F}_\text{ch}(Q^2)$.

The Wilson coefficients in~\cref{effective_Lagrangian_quarks} are calculated via a detailed matching performed at the scale $\mu=M_Z$ (the Higgs fields, top quark, $W$ and $Z$ gauge bosons are integrated out at this step), and then evolved down to low energy scales via a renormalization group analysis, decoupling quark flavors ($b$) as they become heavy (see~\cite{Hill:2019xqk} for a comprehensive description). By performing the matching at $\mu=M_Z$ and running the couplings down to $\mu=2$ GeV, loop corrections from all heavy particles in the SM are systematically included. Four quarks $u$, $d$, $s$, and $c$, are treated as dynamical degrees of freedom. The effects of $u$, $d$, and $s$ are encoded via a nonperturbative charge-charge and charge-isospin current-current correlators, $\hat{\Pi}^{(3)}_{\gamma\gamma} $ and $\hat{\Pi}^{(3)}_{3\gamma} $, whereas the $c$ quark is included perturbatively. Charged leptons are kept as propagating degrees of freedom and their influence on the left- and right-handed couplings is included explicitly via a loop expansion. For heavy particles (tau leptons, the charm quark, and hadronic loops), the momentum transfer is approximated as $Q^2=0$. For light leptons ($e$ and $\mu$), the full $Q^2$ dependence is essential.
\begin{figure}
          \centering
          \includegraphics[height=0.23\textwidth]{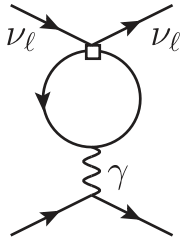}              
          \caption{Closed fermion loop contribution to neutral-current neutrino scattering. This diagram includes both charged and neutral currents \emph{c.f.}~\cref{WZ-diags}. Heavy fermion loops, with masses above the renormalization scale, are included implicitly in the four-fermion Wilson coefficients. The effects of the light fermion loops (i.e.\ $e,\mu,\tau$ and $u,d,s,c$) can be captured by taking tree-level expressions and making the replacement $c^i_{\rm L,R}\rightarrow \tilde{c}^{i,\nu_\ell}_{\rm L,R}$ as described in~\cref{eq:radiative_correction-1}.  \label{fig:closed_loop_ET}}
\end{figure}

In calculations of \cevns cross sections, loop-level effects from light degrees of freedom can be conveniently captured by taking tree-level expressions 
and replacing $c^i_{\rm L,R}\rightarrow \tilde{c}^{i,\nu_\ell}_{\rm L,R}(Q^2)$ everywhere.\
The tilded couplings, $\tilde{c}^{i,\nu_\ell}_{\rm L,R}(Q^2)$, are $Q^2$ dependent as they include the effects of dynamical lepton and quark-mediated loops from~\cref{fig:closed_loop_ET}. Note that due to this $Q^2$ dependence, tilded couplings are not proper Wilson coefficients. They are given explicitly by
\begin{align}
    \tilde{c}^{i,\nu_\ell}_\mathrm{L,R} &= c^i_\mathrm{L,R} +  \frac{\alpha}{\pi}  \frac{\mathrm{G}_\mathrm{F}}{\sqrt{2}} Q_i\delta^{\mathrm{QCD}} +  \frac{\alpha}{\pi}  \frac{\mathrm{G}_\mathrm{F}}{\sqrt{2}} Q_i\delta^{\nu_\ell}, \label{eq:radiative_correction-1} \\
    \delta^{\nu_\ell} &=  \frac{c_{\mathrm{L}}^{\nu_{\ell} e} + c_{\mathrm{R}}^{\nu_{\ell} e} }{\sqrt{2}\mathrm{G}_\mathrm{F}} \mathrm{\Pi} \left( Q^2,m_e; \mu \right) +\frac{c_{\mathrm{L}}^{\nu_{\ell} \mu} + c_{\mathrm{R}}^{\nu_{\ell} \mu} }{\sqrt{2}\mathrm{G}_\mathrm{F}} \mathrm{\Pi} \left( Q^2,m_\mu; \mu \right)+ \frac{c_{\mathrm{L}}^{\nu_{\ell} \tau} + c_{\mathrm{R}}^{\nu_{\ell} \tau} }{\sqrt{2}\mathrm{G}_\mathrm{F}} \mathrm{\Pi} \left( 0,m_\tau; \mu \right), \label{eq:radiative_correction-2}\\
    \delta^{\mathrm{QCD}}  &= 4  \left( \hat{\Pi}_{\gamma \gamma}^{(3)}(0; \mu) \sin^2 \theta_\mathrm{W}  - \frac{1}{2} \hat{\Pi}_{3 \gamma}^{(3)}(0; \mu) \right)  - N_c Q_c\frac{c_{\mathrm{L}}^{c} + c_{\mathrm{R}}^{c} }{\sqrt{2}\mathrm{G}_\mathrm{F}} \mathrm{\Pi} \left( 0,m_c; \mu \right),\label{eq:radiative_correction-3}
\end{align}
where $Q_i$ is the electric charge of the particle in units of the proton charge, $N_c = 3$ is the number of colors in QCD, $\hat{\Pi}_{3 \gamma}^{(3)}(0; \mu)$ is a nonperturbative current-current charge-isospin correlator in the theory with $n_f=3$ quark flavors estimated as $\hat{\Pi}_{3 \gamma}^{(3)} = (1\pm 0.2) \hat{\Pi}_{\gamma \gamma}^{(3)}$~\cite{Tomalak:2019ibg} with the charge-charge correlator evaluated from the experimental data on hadron production $\hat{\Pi}_{\gamma \gamma}^{(3)}(0; \mu = 2~\mathrm{GeV}) = 3.597(21)$~\cite{Erler:1998sy,Erler:2004in,Erler:2017knj}. For numerical estimates, we take $\mathrm{G}_\mathrm{F}=1.1663787\times 10^{-5}\,{\rm GeV}^{-2},~\sin^2\theta_\mathrm{W} = 0.23112,~\alpha^{-1}(2\,{\rm GeV}) = 133.309$ and the charm quark mass $\hat{m}_c(2\,{\rm GeV}) = 1.096~\mathrm{GeV}$,\footnote{We use the value of the Weinberg angle at the scale $\mu = M_Z$ since it enters our corrections in~\cref{eq:radiative_correction-3} with a factor $\alpha$ and any effects of running are more than order of magnitude below the size of hadronic errors.} masses of charged leptons from PDG~\cite{Tanabashi:2018oca} and coupling constants from Table 1 of~\cite{Tomalak:2019ibg}. For convenience, we present the effective couplings in Table~\ref{results_couplings_Running} and compare them to the naive tree-level determination.
\begin{table}[tbp]
\footnotesize
\centering
\begin{tabular}{|c|c|c|c|c|c|c|c|c|}   
\hline          
$ c^{\nu_\ell \ell'}_\mathrm{L},~\ell = \ell'$ & $ c^{\nu_\ell \ell'}_\mathrm{L},~\ell \neq \ell'$ & $  c^{\nu_\ell \ell'}_\mathrm{R}$ & $ c^{u}_\mathrm{L}$ & $ c^{u}_\mathrm{R}$ & $ c^{d}_\mathrm{L}$ & $ c^{d}_\mathrm{R}$ \\
\hline
  $2.39818(33)$ &$-0.90084(32)$ &$0.76911(60)$ &$1.14065(13)$ &$-0.51173(38)$ &$-1.41478(12)$ &$0.25617(20)$  \\
\hline
\hline 
\hline
 $2.412$ & $-0.887$ &$0.763$ &  $1.141$ & $-0.508$ & $-1.395$  & $0.254$   \\
 \hline
\end{tabular}
\caption{\textbf{Top row:} Effective couplings (in units $10^{-5}~\mathrm{GeV^{-2}}$) in the Fermi theory of neutrino-fermion scattering with four quark flavors within the $\overline{\text{MS}}$ renormalization scheme at the scale $\mu = 2~\mathrm{GeV}$. The error due to the uncertainty of Standard Model parameters is added in quadrature to a perturbative error of matching. For illustration, we have included the tree-level couplings. Coefficients are determined at the scale $\mu=M_Z$ via a matching calculation and then run down to $\mu=2$ GeV via a renormalization group analysis. For a more detailed discussion see~\cite{Hill:2019xqk}. \textbf{Bottom row:} Tree-level expressions for the same quantities are quoted to three decimal places. Notice that some Wilson coefficients receive $\mathrm{O}(1\%)$ corrections. Tree level expressions are defined as $c_\mathrm{L}^{\nu_\ell \ell^\prime} = 2\sqrt{2} \mathrm{G}_\mathrm{F} (\sin^2\theta_\mathrm{W} - 1/2 + \delta_{\ell\ell'})$, $c_\mathrm{R}^{\nu_\ell \ell^\prime} = 2\sqrt{2} \mathrm{G}_\mathrm{F}\sin^2\theta_\mathrm{W} $, $c_\mathrm{R}^{ q}= 2\sqrt{2} \mathrm{G}_\mathrm{F} (-Q_q \sin^2\theta_\mathrm{W})$, and $c_\mathrm{L}^{ q}=2\sqrt{2} \mathrm{G}_\mathrm{F} (T_q^3-Q_q \sin^2\theta_\mathrm{W})$ with the quark charge $Q_q$ and isospin $T_q^3$, they are evaluated using $\mathrm{G}_\mathrm{F}=1.1663787\times 10^{-5}\,{\rm GeV}^{-2}$ and $\sin^2\theta_\mathrm{W} = 0.23112$. 
\label{results_couplings_Running}}
\end{table}

Also appearing in~\cref{eq:radiative_correction-2,eq:radiative_correction-3} is the vacuum polarization function, $\Pi(Q^2,m_f;\mu)$ (familiar from QED) which is generated by closed fermion loops coupling to photons shown in~\cref{fig:closed_loop_ET}. At the renormalization scale $\mu$ in the $\overline{\mathrm{MS}}$ scheme,\footnote{We quote the result for an arbitrary renormalization scale. However, $\mu=2$ GeV should be used in conjunction with the couplings in~\cref{results_couplings_Running}.} the vacuum polarization function $\mathrm{\Pi} \left( Q^2,m_f; \mu \right)$ is given by~\cite{Pauli:1936zz,Feynman:1949zx,Tsai:1960zz,Vanderhaeghen:2000ws,Heller:2019dyv}
\begin{align}
    \mathrm{\Pi} \left( Q^2,m_f; \mu \right) &= \frac{1}{3} \ln \frac{\mu^2}{m_f^2} + \frac{5}{9}  - \frac{4m_f^2}{3 Q^2} + \frac{1}{3} \left( 1 - \frac{2m_f^2}{Q^2} \right)  \sqrt{1+\frac{4m_f^2}{Q^2}} \ln \frac{  \sqrt{1+\frac{4m_f^2}{Q^2}} - 1 }{\sqrt{1+\frac{4m_f^2}{Q^2}} + 1}\nonumber\\
    &=\begin{cases}
    \frac{1}{3} \ln \frac{\mu ^2}{m_f^2}-\frac{Q^2}{15 m_f^2} +\mathrm{O}\left(Q^4/m_f^4\right), & m_f^2 \gg Q^2;\\
    \frac{1}{3} \ln \frac{\mu ^2}{Q^2}+ \frac{5}{9} +\mathrm{O}\left(m_f^2/Q^2\right),
    & m_f^2 \ll Q^2~.
    \end{cases}
    \label{vacuum_polarization}
\end{align}
For $m_f=m_e$ (in particular) and $Q=1-50$ MeV, there is no sense in which a small $Q^2$ approximation is justified, and the full $Q^2$ dependence must be retained. Tau leptons are sufficiently heavy that the $Q^2\rightarrow 0$ limit can be taken safely for \cevns kinematics. 

We also include $\alpha_s$ contributions to the charm-quark closed loop; for analytical expressions see~\cite{Djouadi:1987gn,Djouadi:1987di,Kniehl:1989yc,Fanchiotti:1992tu,Tomalak:2019ibg} and references therein. It is instructive to note that radiative corrections of vector type do not change the axial part of the nucleon current (see~\cref{neutral_current_form_factor}). Consequently, all radiative corrections (besides the nuclear vertex correction) are described by substitutions of~\cref{eq:radiative_correction-1} for nuclei of arbitrary spin. The factorizable part of the vertex correction and photon bremsstrahlung from the heavy nucleus can be safely neglected. 

For nuclei of nonzero spin, QED vertex corrections introduce higher electromagnetic moments of order $\alpha$. These corrections are beyond the scope of this work, but can be included in the definition of electromagnetic response functions and can therefore be folded into any empirical determination of nuclear form factors. All residual effects scale as $Q^2/M_\mathrm{A}^2$ and are negligible for heavy nuclei. We therefore expect that the results of this paper can be extended to higher spin nuclei in a relatively straightforward manner. 

\subsection{Nucleon form factors in \cevns\label{nucleons}}
Embedding quarks into nucleons, the matrix element of the quark current is expressed in terms of Sachs electric $\mathrm{G}^q_\mathrm{E}$ and magnetic $\mathrm{G}^q_\mathrm{M}$ isovector, axial $\mathrm{F}^q_\mathrm{A}$, and pseudoscalar $\mathrm{F}^q_\mathrm{P}$ form factors for individual quarks as
\ber \label{neutral_current_form_factor}
\Gamma_\mu (Q^2) &=&  \langle N(p^\prime) | \sum_{q} \bar{q} \gamma_\mu (c_\mathrm{L}^{q} \mathrm{P}_\mathrm{L} + c_\mathrm{R}^{q} \mathrm{P}_\mathrm{R}) q  | N(p) \rangle  \nonumber \\
&=& \sum_{q} \frac{c_\mathrm{L}^{q} +c_\mathrm{R}^{q} }{2} \bar{N}\! \left[ 
\gamma_\mu \mathrm{G}_M^{q, \mathrm{N}}(Q^2) - \frac{p_\mu + p'_\mu}{2M_\mathrm{N}} \frac{\mathrm{G}_\mathrm{M}^{q, \mathrm{N}} (Q^2) - \mathrm{G}_\mathrm{E}^{q, \mathrm{N}} (Q^2)}{1+\tau_\mathrm{N}} \right]\! N \nonumber \\
&~&\hspace{60pt}+ \sum_{q} \frac{c_\mathrm{L}^{q} - c_\mathrm{R}^{q} }{2} \bar{N}\! \left[ 
\gamma_\mu \gamma_5 \mathrm{F}^{q, \mathrm{N}}_\mathrm{A}(Q^2) + \frac{l_\mu}{M_\mathrm{N}} \gamma_5 \mathrm{F}^{q, \mathrm{N}}_\mathrm{P}(Q^2) \right]\! N \,,
\eer
with $ l = p' - p$, $Q^2 \equiv -l^2 = - \left( p - p'\right)^2$ and $\tau_\mathrm{N}=Q^2/(4M_\mathrm{N}^2)$, where $M_\mathrm{N}$ is the nucleon mass. We concentrate on the vector part contributing to CE$\nu$NS cross section on a spin-0 target. Assuming isospin symmetry, the proton ($p$) and neutron ($n$) form factors can be expressed in terms of the quark contributions to the proton form factor $\mathrm{G}^{q,p}$ as\footnote{The neglected relative difference in nucleon masses contributes a correction below permille level. Isospin-breaking effects are expected in constituent quark model to be at the similar level~\cite{Behrends:1960nf,Dmitrasinovic:1995jt,Miller:1997ya,Lewis:2007qxa}. According to ChPT-based calculation of~\cite{Kubis:2006cy}, they can be much larger. This question requires a further theoretical investigation on the lattice.}
\begin{align}
    \mathrm{G}_\mathrm{E,M}^p &=\frac{2}{3} \mathrm{G}_\mathrm{E,M}^{u,p} -\frac{1}{3} \mathrm{G}_\mathrm{E,M}^{d,p} - \frac{1}{3} \mathrm{G}_\mathrm{E,M}^{s,p} ,\label{proton-FF} \\
    \mathrm{G}_\mathrm{E,M}^n & =  \frac{2}{3} \mathrm{G}_\mathrm{E,M}^{d,p} - \frac{1}{3} \mathrm{G}_\mathrm{E,M}^{u,p} - \frac{1}{3} \mathrm{G}_\mathrm{E,M}^{s,p} ~ \label{neutron-FF}.
\end{align}
For applications to CE$\nu$NS at low momentum transfer, we need only the normalization of electromagnetic form factors and the first term in the Maclaurin series which provides a conventional definition of the associated radius $r^{i}_\mathrm{E,M}$
\beq
    \mathrm{G}^{i}_\mathrm{E,M} \left( Q^2 \right) = \mathrm{G}^{i}_\mathrm{E,M} \left( 0 \right) \left[1 - \frac{ \left(r^i_\mathrm{E,M}\right)^2}{6}Q^2 + \mathrm{O} \left(Q^4 \right) \right] \,.  \label{radius_definition}
\eeq 
More specifically, we need mainly the normalization of the electric form factor, given explicitly by $\mathrm{G}_\mathrm{E}^{u,p}(0)=2$, $\mathrm{G}_\mathrm{E}^{d,p}(0)=1$, and $\mathrm{G}_\mathrm{E}^{s,p}(0)=0$, and electric charge radii. The most precise determination of the proton electric charge radius is obtained from the muonic hydrogen Lamb shift~\cite{Pohl:2010zza,Antognini:1900ns}:
\beq
    r_\mathrm{E}^p = 0.84087(39) \text{ fm} \,.  \label{radius_proton}
\eeq
The neutron electric charge radius is measured precisely scattering neutrons on heavy targets~\cite{Kopecky:1995zz, Kopecky:1997rw}:
\beq
  \langle r^2 \rangle^n_\mathrm{E} = -0.1161(22) \text { fm}^2 \,. \label{radius_neutron}
\eeq
For the strange electric charge radius and magnetic moment, calculations from lattice QCD~\cite{Shanahan:2014tja,Green:2015wqa,Sufian:2016pex,Djukanovic:2019jtp,Alexandrou:2019olr} have recently appeared. We choose the result with a conservative error estimate from~\cite{Djukanovic:2019jtp}:
\beq
  \langle r^2 \rangle^s_\mathrm{E} = -0.0046(18) \text { fm}^2, \qquad \mu^s = -0.020(13)  \,.\label{radius_strange}
\eeq
The strange quark contributes negligibly in the kinematics of neutrinos with energy below $100~\mathrm{MeV}$. Removing the strange quark from proton constituents does not change any numbers in this manuscript within significant digits.

\subsection{Weak and charge nuclear form factors in \cevns \label{nucleus}}

Form factors entering the cross section expressions can be defined as the product of nucleon-level form factors with point-proton and point-neutron distributions inside the nucleus $f_p \left(Q^2\right)$ and $f_n \left(Q^2\right)$. Using the isospin-decomposed nucleon form factors of~\cref{proton-FF,neutron-FF} yields the following definition for the renormalization scale-dependent weak and charge form factors~\cite{Serot:1978vj,Walecka:1995mi,Horowitz:2012we,Horowitz:2012tj,Hoferichter:2020osn}\footnote{The normalization of form factors $f_p$ and $f_n$ is fixed by the number of protons ($Z$) and neutrons ($N$) inside the nucleus as $f_p \left(0\right) = Z$ and $f_n \left(0\right) = N$.}
\begin{align} \label{eq:form_factor_weak}
    \mathrm{F}_\mathrm{W} &=\left( \frac{c_\mathrm{L}^{u} +c_\mathrm{R}^{u} }{\sqrt{2} \mathrm{G}_\mathrm{F}} \mathrm{G}_\mathrm{E}^{u,p} + \frac{c_\mathrm{L}^{d} +c_\mathrm{R}^{d} }{\sqrt{2} \mathrm{G}_\mathrm{F}}  \mathrm{G}_\mathrm{E}^{d,p} \right) f_p + (p \leftrightarrow n), \\
    \mathrm{F}_\mathrm{ch} &= \mathrm{G}_\mathrm{E}^p f_p + (p \leftrightarrow n)~,\label{eq:form_factor_charge}
\end{align}
where we have omitted small Darwin-Foldy (DF) and spin-orbit terms~\cite{Hoferichter:2020osn} that contribute below the uncertainty of hadronic corrections in the kinematic region of CE$\nu$NS experiments from pion decay at rest ($\pi$DAR) sources. In our detailed calculations, we include these effects and find them to be negligible (contributing at the sub-permille level). 

 
As discussed above, heavy physics mediated loops are included implicitly in the definition of $c_{\rm L}^i$ and $c_{\rm R}^i$. This leads to a modification of the weak form factor of the proton and neutron relative to their tree-level values. This definition is somewhat conventional since certain electromagnetic corrections, such as those proportional to $\delta^\text{QCD}$, could reasonably be shuffled into the definition of the weak form factor. Our convention separates electromagnetic corrections mediated by light degrees of freedom such that the definition of the weak charge of the proton and neutron is given, respectively, by
\begin{align}
    Q_\text{W}^p \left( \mu = 2~\mathrm{GeV} \right) &= 2 \frac{c_\mathrm{L}^{u} +c_\mathrm{R}^{u} }{\sqrt{2} \mathrm{G}_\mathrm{F}} + \frac{c_\mathrm{L}^{d} +c_\mathrm{R}^{d} }{\sqrt{2} \mathrm{G}_\mathrm{F}} = 0.06015(53) , \\
    Q_\text{W}^n  &=
    2 \frac{c_\mathrm{L}^{d} +c_\mathrm{R}^{d} }{\sqrt{2} \mathrm{G}_\mathrm{F}} + \frac{c_\mathrm{L}^{u} +c_\mathrm{R}^{u} }{\sqrt{2} \mathrm{G}_\mathrm{F}} = -1.02352(25),
\end{align}
and can be compared to the tree-level values $Q_\text{W}^p = 0.0751$ and $Q_\text{W}^n = -1$.\footnote{At leading order, the weak charge of the nucleus is given approximately by the number of neutrons $N$ taken with opposite sign.} The renormalization scale dependence in the proton weak charge is inherited from the Wilson coefficients.

Alternatively, one can define the scale-independent but process-dependent weak charge of the nucleus entering the cross section at $Q^2 = 0 $ as
\begin{align}
Q^{\nu_\ell}_\mathrm{W} &= \mathrm{F}_\mathrm{W} \left( 0 \right) + \frac{\alpha}{\pi} \left( \delta^{\nu_\ell} + \delta^\mathrm{QCD}\right) \mathrm{F}_\mathrm{ch} \left(0 \right) = Z Q_\text{W}^{p,\nu_\ell} + N Q_\text{W}^n, \label{eq:charge_difference} \\ 
Q_\text{W}^{p,\nu_\ell} &= 2 \frac{c_\mathrm{L}^{u} \left( \mu \right) +c_\mathrm{R}^{u} \left( \mu \right) }{\sqrt{2} \mathrm{G}_\mathrm{F}} + \frac{c_\mathrm{L}^{d} \left( \mu \right) +c_\mathrm{R}^{d} \left( \mu \right) }{\sqrt{2} \mathrm{G}_\mathrm{F}} + \frac{\alpha}{3\pi} \sum \limits_{\ell'=e,\mu,\tau} \frac{c_{\mathrm{L}}^{\nu_{\ell} \ell'}\left( \mu \right)+ c_{\mathrm{R}}^{\nu_{\ell} \ell'} \left( \mu \right)}{\sqrt{2}\mathrm{G}_\mathrm{F}} \ln \frac{\mu^2}{m^2_{\ell'}}
\nonumber \\
& + \frac{4 \alpha}{\pi} \left( \hat{\Pi}_{\gamma \gamma}^{(3)}(0; \mu) \sin^2 \theta_\mathrm{W}  - \frac{1}{2} \hat{\Pi}_{3 \gamma}^{(3)}(0; \mu) -\frac{c_{\mathrm{L}}^{c}\left( \mu \right) + c_{\mathrm{R}}^{c}\left( \mu \right) }{2\sqrt{2}\mathrm{G}_\mathrm{F}} \mathrm{\Pi} \left( 0,m_c; \mu \right) \right).
\label{eq:charge_difference2}
\end{align}
This illustrates that the definition of a weak nuclear charge is not unique at one-loop order and beyond. Rather, these quantities depend on choice of convention, on the renormalization scheme and on the process in general. 

One should not compare directly the weak form factor and the weak charge in our definition to commonly used values in parity-violating electron scattering (PVES)~\cite{Tanabashi:2018oca}. The Wilson coefficients that will be embedded inside nucleons are process dependent, i.e.\ electron-quark and neutrino-quark couplings differ after accounting for heavy-physics mediated one-loop corrections. Moreover, one-loop radiative corrections depend on the process even within the low-energy effective theory. For example, in PVES electrons can exchange a photon with the nucleus leading to a $\gamma-Z$ box diagram at NLO. This cannot occur in \cevns because neutrinos do not couple directly to photons. A comparison of the weak nuclear charge defined in this work to conventions in the PVES literature will require a proper account for one-loop running of electron-quark Wilson coefficients from the electroweak scale down to $\mu=2$ GeV and reliable treatment of radiative corrections in PVES. 

As for neutrino charge radii $r^2_{\nu_\ell}$~\cite{Hill:2019xqk}, QCD and uncertain hadronic contributions cancel in flavor differences of charges in~\cref{eq:charge_difference,eq:charge_difference2}. One can precisely predict these differences up to $\mathrm{O} \left( \alpha^3 \right)$ in terms of the charge $e_0$ and electromagnetic coupling constant $\alpha_0 $ in the Thomson limit,~$\alpha_0^{-1}= 137.035 999 084 (21)$~\cite{Tanabashi:2018oca}, as
\begin{align}
     Q^{\nu_\ell}_\mathrm{W} - Q^{\nu_{\ell'}}_\mathrm{W} &=  \frac{Z e^2_0 \left( r^2_{\nu_{\ell'}} - r^2_{\nu_{\ell}} \right)}{3 \sqrt{2} \mathrm{G}_\mathrm{F}} = \frac{2 Z \alpha_0}{3\pi} \ln \frac{m^2_{\ell'}}{m^2_\ell} \left( 1 + \frac{3 \alpha \left( \mu \right)}{4 \pi} + ... \right), \\
     Q^{\nu_e}_\mathrm{W} - Q^{\nu_\mu}_\mathrm{W} &= 0.01654 Z ,\\
     Q^{\nu_\mu}_\mathrm{W} - Q^{\nu_\tau}_\mathrm{W} &= 0.00876 Z,
\end{align}
while the absolute scale of weak charge is determined by the proton $Q^{p,\nu_e}_\mathrm{W} = 0.0747(34)$ and scale- and flavor-independent neutron weak charge $ Q_\text{W}^n  = -1.02352(25)$.\footnote{Our flavor differences are in agreement with~\cite{Erler:2013xha,Cadeddu:2020lky} while central values for weak charges differ due to distinct matching and treatment of hadronic contributions in~\cite{Erler:2013xha,Cadeddu:2020lky} vs~\cite{Hill:2019xqk}.}

This provides a microscopic definition of nuclear form factors in terms of the Wilson coefficients in the four-fermion Lagrangian presented above, assuming a conventional picture of nuclear physics. Trusting such a microscopic picture relies on a top-down approach to nuclear physics, and one may worry that complicated effects related to such a many-body strongly interacting system are not well understood. One famous example is the EMC effect~\cite{Aubert:1983xm,Hen:2016kwk,Schmookler:2019nvf}, where quark parton distribution functions inside nuclei were found to differ from those of nucleons in a vacuum. This effect is still not understood. Nevertheless, at momentum transfers of $\mathrm{O}(50~\text{MeV})$ and below, one does not expect these effects to be important.

An alternative to this is the bottom-up approach based on the extraction of relevant form factors, $\mathrm{F}_\text{ch}(Q^2)$ and $\mathrm{F}_\text{W}(Q^2)$, directly from experiments. In fact, $\mathrm{F}_\text{ch}(Q^2)$ is already well determined with high precision for a number of nuclei~\cite{Angeli:2013epw, DeJager:1987qc}, including $^{40}$Ar~\cite{Ottermann:1982kr}, through decades of elastic electron scattering experiments. For instance, the charge radius of $^{208}$Pb extracted from experiment is known to about 0.02\% precision~\cite{Angeli:2013epw}. Similarly, $\mathrm{F}_\text{W}(Q^2)$ can be experimentally determined in a clean model-independent way from electroweak probes such as \cevns and PVES experiments~\cite{Donnelly:1989qs}. 

A precise measurement of \cevns cross section on a particular nucleus can be used to extract $\mathrm{F}_\text{W}(Q^2)$ of that nucleus, using~\cref{cevns-cross-section}. Note however that, while $\delta^{\nu_{\ell}}$ can be calculated and $\mathrm{F}_\text{ch}(Q^2)$ can be measured with electron-nucleus scattering, the precision on the determination of $\mathrm{F}_\text{W}(Q^2)$ will be limited by hadronic uncertainties stemming from $\delta^{\text{QCD}}$. The latter nonperturbative object can be in principle constrained by lattice calculations or performing measurements on nuclei with different numbers of protons and neutrons, $Z$ and $N$ respectively. Another possible way of measuring the weak form factor of the nucleus is via PVES experiments, albeit subject to the difficulties outlined above relating to process-dependent shifts in Wilson coefficients and radiative corrections. To convey the basic idea, we ignore these subtleties and work at leading order in the following discussion.
The key experimental observable in the elastic scattering of longitudinally polarized electrons from the unpolarized spin-0 nucleus is the parity-violating asymmetry $\mathrm{A}_\mathrm{PV}$. 
The parity-violating asymmetry arises from the interference of $\gamma$-mediated and $Z$-mediated scattering diagrams. 
The asymmetry $\mathrm{A}_\mathrm{PV}$ is determined from the fractional difference in cross sections between the scattering of positive and negative helicity electrons, that is, $\mathrm{A}_\mathrm{PV} = (\sigma_+-\sigma_-)/(\sigma_++\sigma_-)$, where $\pm$ refers to the polarization of the electron.
This is similar to the parity violation asymmetry in M\"{o}ller scattering experiments. 
In the Born approximation at low momentum transfer, $\mathrm{A}_\mathrm{PV}$ is proportional to the ratio of the weak to the charge form factors of the nucleus:\footnote{PVES is sensitive to Wilson coefficients of electron-quark interaction and is subject to radiative corrections distinct to the \cevns process.}
\begin{align}\label{eq:PVES}
    \mathrm{A}_\mathrm{PV} \approx \frac{\mathrm{G}_\mathrm{F} Q^2}{4 \sqrt{2} \pi \alpha}~ \frac{\mathrm{F}_\text{W}(Q^2)}{\mathrm{F}_\text{ch}(Q^2)},
\end{align}
where form factors are normalized to the nucleus' weak and electric charges, that is, $ Q_\text{W} \equiv \mathrm{F}_\text{W}(Q^2 = 0) $ and $Z \equiv \mathrm{F}_\text{ch}(Q^2 = 0)$. For a given nucleus, if $\mathrm{F}_\text{ch}(Q^2)$ is already known from elastic electron scattering experiment, one can extract $\mathrm{F}_\text{W}(Q^2)$ from measured $\mathrm{A}_\mathrm{PV}$ in~\cref{eq:PVES} at the momentum transfer of the experiment after accounting for radiative corrections and Coulomb distortion effects not considered in the Born approximation~\cite{Horowitz:1999fk}. Coulomb distortions can be theoretically calculated by solving the Dirac equation for an electron moving in a nuclear potential~\cite{Yennie:1954zz,Yennie:1965zz,Czyz:1963ads,Kim:1996ua,Kim:2001sq} and are relatively well understood~\cite{Horowitz:1998vv}.

In fact, the PREX experiment at JLab has done such measurement and provided the first model-independent determination of the weak form factor of $^{208}$Pb, $\mathrm{F}_\text{W}(\langle Q^2\rangle ) = 0.204\pm0.028$ at the average momentum transfer of the experiment $\langle Q^2\rangle \approx 8800~\mathrm{MeV}^2$~\cite{Abrahamyan:2012gp, Horowitz:2012tj}. The PREX-II experiment is currently underway and is expected to improve the precision of the $^{208}$Pb form factor measured by PREX. The CREX experiment is planned to measure the weak form factor of $^{48}$Ca~\cite{Kumar:2020ejz}. Future facilities such as the MESA facility in Mainz envisioned to start operations in a few years will also be suited for high-precision parity-violating experiments~\cite{Becker:2018ggl}.

It is worth noting that \cevns can be used to probe the weak form factor only at low momentum transfers where the process remains coherent, but accesses a continuum of four-momentum transfers. In contrast, PVES experiments are usually carried out at a single value of the momentum transfer at a time. A combination of measurements from these two independent and complementary scattering techniques is ideal since systematic uncertainties are largely uncorrelated. This will then provide an empirical extraction of a nucleus' weak form factor in clean and model-independent fashion. 


\vfill 
\subsection{Cross sections and uncertainties for a monoenergetic source \label{error-budget} }

In this section, we evaluate CE$\nu$NS cross sections and provide a complete error budget for the case of $^{40}$Ar nucleus.\footnote{See~\cite{Cheoun:2011zza,Capozzi:2018dat,Gayer:2019eed} for a recent discussion of neutrino neutral-current cross sections on $^{40}$Ar producing an excited state of the nucleus.} We focus on three benchmark energies relevant for a \cevns experiment whose neutrino flux is sourced by pion decay at rest ($\pi$DAR). 
We take $E_\nu=$ 10, 30, and 50 MeV. 
A proper treatment of flux-averaged cross sections in a realistic experimental setup would require a precise prediction of the daughter neutrinos from both pion and muon decay. 
This necessarily involves an understanding of decay in flight (DIF) contamination and a more precise theoretical prediction of both the pion's and muon's decay spectrum including radiative corrections; both effects need to be understood at the percent or even permille level. 
The former issue is experiment specific and must be addressed with Monte Carlo simulations, while the latter is an interesting and important theoretical problem that we leave to future work. We discuss some of these issues qualitatively in~\cref{flux-avg}. 

\begin{figure}                            \centering
    \includegraphics[height=0.226\textwidth]{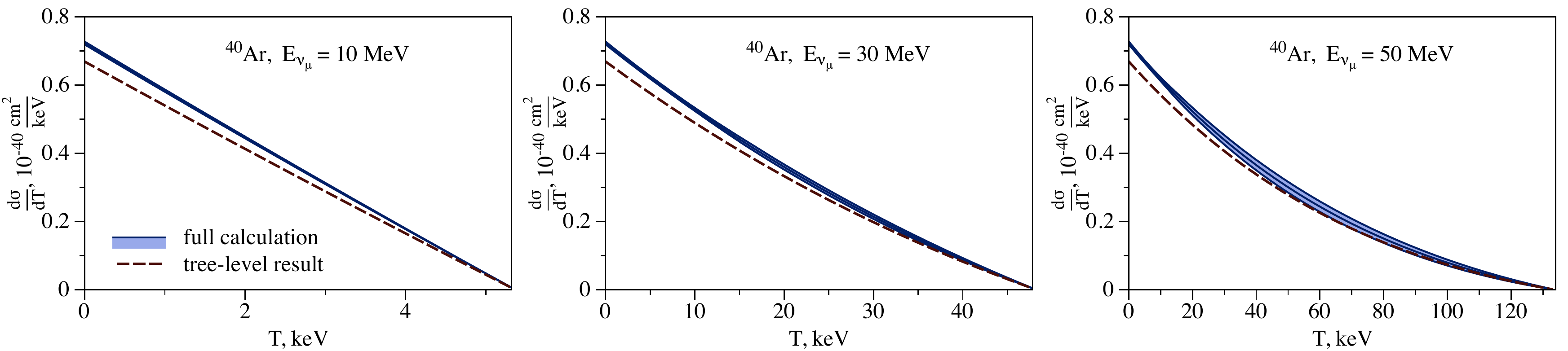} 
    \caption{CE$\nu_\mu$NS cross section on$~^{40} \mathrm{Ar}$ target (blue band) compared to the tree-level prediction (red dashed curve) as a function of the recoil nucleus energy for the incoming neutrino energies $E_{\nu_\mu} = 10,~30$, and $50~\mathrm{MeV}$. The error bands shown here include the full error budget outlined in~\cref{error-budget} and summarized in~\cref{tab:errors_Argon}. \label{fig:CEvNS_Argon}} 
\end{figure}

In what follows, we provide an estimated error budget accounting for uncertainties stemming from a variety of sources. Our treatment is meant to be conservative and exhaustive. We include the following sources of uncertainties: 
\begin{itemize}
    \item \textbf{Nuclear level}: For the central values, we average over eight available nuclear calculations for $f_p$ and $f_n$~\cite{Yang:2019pbx,Payne:2019wvy,Hoferichter:2020osn,VanDessel:2020epd}. At small momentum transfer, we estimate the theoretical uncertainty using the small $Q^2$ expansion in terms of point-nucleon radii $R_p = 3.338\pm0.003~\mathrm{fm},~R_n = 3.406\pm0.046~\mathrm{fm}$~\cite{Angeli:2013epw,Payne:2019wvy,Barbieri:2019ual},\footnote{Shell-model~\cite{Hoferichter:2020osn} and density functional theory calculations predict larger values for the neutron skin $R_n - R_p \lesssim 0.11~\mathrm{fm}$. Conservatively increasing error of $R_n$ to $0.065~\mathrm{fm}$ increases the nuclear error in~\cref{tab:errors_Argon} at $E_\nu = 10~\mathrm{MeV}$ as $0.04 \to 0.06$ and in~\cref{tab:errors_Argon_difference} at $E_\nu = 10~(30,50)~\mathrm{MeV}$ as $0.82 \to 0.83~(0.21 \to 0.28,~0.02 \to 0.03)$ respectively. This modified error estimate increases the error of the flavor asymmetry in~\cref{tab:errors_Argon_difference} at energy $E_\nu = 50~\mathrm{MeV}$ by the last significant digit, while the total error in~\cref{tab:errors_Argon_difference} at energy $E_\nu = 30~\mathrm{MeV}$ increases as $0.35 \to 0.40$.} i.e. $f_p= Z \left(1 - {Q^2 R_p^2}/{6} \right), f_n= N \left(1 - {Q^2 R_n^2}/{6} \right)$, as the error propagated from point-proton and point-neutron radii added in quadrature to the error of next terms in the $Q^2$ expansion estimated as $ \left[\sigma(R_p, R_n) - \sigma (R_p, R_n =0)\right]^2/\sigma(R_p, R_n)$. At larger momentum transfer, we take the largest difference between the theoretical calculations of~\cite{Patton:2012jr,Yang:2019pbx,Payne:2019wvy,Hoferichter:2020osn,VanDessel:2020epd}. At intermediate region, we take the minimum of these two estimates. 
    
    \item \textbf{Nucleon level:} We exploit a $Q^2$ expansion for the error estimate at the nucleon level. We add in quadrature propagated errors of charge radii of~\cref{radius_proton,radius_strange} to the estimate of neglected terms in $Q^2$ expansion:
    \ber
    \left[\sigma(r_\mathrm{E}^p, \langle r^2\rangle_\mathrm{E}^n, \langle r^2\rangle_\mathrm{E}^s) - \sigma (r_\mathrm{E}^p, \langle r^2\rangle_\mathrm{E}^n, \langle r^2\rangle_\mathrm{E}^s = 0)\right]^2/\sigma (r_\mathrm{E}^p, \langle r^2\rangle_\mathrm{E}^n, \langle r^2\rangle_\mathrm{E}^s).
    \eer
    We also include an uncertainty due to the isospin symmetry breaking given by the relative difference of proton and neutron masses multiplying the radii central values.
    
    \item \textbf{Hadronic contributions:} Following~\cite{Tomalak:2019ibg}, we take the correlator $\hat{\Pi}_{3\gamma}^{(3)}$ in the approximation of exact $\mathrm{SU}(3)_f$ symmetry when $\hat{\Pi}^{(3)}_{3\gamma} = \hat{\Pi}_{\gamma\gamma}$ with $\hat{\Pi}_{\gamma\gamma}=3.597(21)$~\cite{Erler:1998sy,Erler:2004in,Erler:2017knj} and conservatively assign a 20\% error.
    
    \item \textbf{Wilson coefficients:} We propagate central values and uncertainties of neutrino-quark coupling constants from~\cite{Tomalak:2019ibg,Hill:2019xqk} properly accounting for correlations and threshold matching errors. 
    
    \item \textbf{Perturbative expansion:} We estimate the perturbative error as a difference between our results evaluated at scales $ \mu = \sqrt{2} \mu_0$ and $\mu = \mu_0/\sqrt{2}$ with $\mu_0 = 2$ GeV.

\end{itemize}
To estimate the total uncertainty, we add all of the above errors in quadrature. We present the relative size of uncertainties of total cross section in~\cref{tab:errors_Argon} and for the differential cross sections in~\cref{fig:CEvNS_Argon}. For convenience, we also provide the \cevns cross sections  for the neutrino energy corresponding to the monochromatic neutrino line from pion decay at rest, $E_{\nu} = \frac{m^2_\pi-m^2_\mu}{2 m_\pi}\simeq29.79$~MeV,
\ber
\sigma_{\nu_\mu} = \left( 15.19 \pm 0.25 \right) \cdot 10^{-40}~\mathrm{cm}^2, \qquad \sigma_{\nu_e} = \left( 15.01 \pm 0.24 \right) \cdot 10^{-40}~\mathrm{cm}^2. \label{piDAR_xsections}
\eer

\begin{table}[t]
\centering
\footnotesize
\centering 
\begin{tabular}{|c|c|c|c|c|c|c||c|c|}   
\hline          
$E_\nu,~\mathrm{MeV}$  &  Nuclear & Nucleon  & Hadronic & Quark  & Pert. & Total & $10^{40}\cdot \sigma_{\nu_\mu},\mathrm{cm^2}$  & $10^{40}\cdot \sigma^0_{\nu_\mu},\mathrm{cm^2}$ 
\\
\hline
50  & 4.   &   0.06  & 0.56 & 0.13 & 0.08  & 4.05 & 34.64(1.36) & 32.05
\\
30  &1.5  & 0.014 & 0.56 & 0.13 & 0.03  & 1.65 & 15.37(0.25) & 14.23
\\
10 & 0.04 & 0.001 & 0.56 & 0.13 & 0.004 & 0.58 & 1.91(0.01)& 1.77
\\
\hline
\end{tabular}
\caption{ Contributions to the relative error (in $\%$) of the total CE$\nu_\mu$NS cross section on$~^{40} \mathrm{Ar}$ target for an incident $\nu_\mu$ neutrino energy $E_\nu$. Cross section with radiative corrections $\sigma_{\nu_\mu}$ is compared to the tree-level result $\sigma^0_{\nu_\mu}$. 
\label{tab:errors_Argon} }
\end{table}

As we can see from~\cref{tab:errors_Argon}, at higher energies the main source of uncertainty for the \cevns cross section comes from nuclear physics.
In fact, this can be traced down to the error of the neutron distribution inside the nucleus. The error stemming from the charge-isospin hadronic correlator, $\hat{\Pi}_{3\gamma}^{(3)}$, is the second largest source of uncertainty for $E_\nu \gtrsim 30$ MeV, and actually dominates over the neutron distribution error at low energies. This is because the point-neutron and point-proton form factors are normalized up to infinite precision at $Q^2 \to 0$ while deviations scale as $Q^2$. Thus, much like in neutrino-electron scattering~\cite{Tomalak:2019ibg}, the charge-isospin correlator is the major theoretical bottleneck for precise predictions. 
\
Fortunately, however, our knowledge of this object can be improved with future lattice QCD studies~\cite{Burger:2015lqa,Ce:2018ziv,Ce:2019imp}, and since it's contribution is $\alpha/\pi$ suppressed for neutrino neutral-current scattering, an uncertainty on the order of 5\% would reduce its contribution to the error budget to the permille level. This would be below the precision necessary for next-generation \cevns experiments to achieve their goals. 
The third largest source of the error is the uncertainty of neutrino-quark couplings from~\cite{Hill:2019xqk}.

One interesting consequence of our calculations is that the dominant source of uncertainty at low momentum transfers $Q^2\lesssim 100~\text{MeV}^2$ is given by microscopic particle physics inputs rather than nuclear modeling. The reason for this is that the point-like treatment of the nucleus becomes such a good approximation at low energies that all nuclear modeling details are subdominant to hadronic uncertainties. This motivates using \cevns as a precision observable at low neutrino energies as we discuss in~\cref{ISODAR}.

\section{Flavor-dependent cross section differences \label{flavor-dep}}
As emphasized above, radiative corrections in \cevns can be split into a flavor-independent and flavor-dependent parts, the latter being induced by charged-lepton loops. Noting that proton and charge form factors induce a negligible uncertainty, and that hadronic and neutron-related uncertainties are flavor universal, one can give a precise description of flavor-dependent effects. 

This simple observation means that the flavor difference can be calculated precisely at $\mathrm{O}(\mathrm{G}_\mathrm{F}^2\alpha)$ subject only to uncertainties related to $\mathrm{F}_\text{W}(Q^2)$. The flavor difference is expressed in terms of weak and charge form factors as 
\begin{equation}\begin{split}\label{eq:flavor-difference}
    \dv{\sigma_{\nu_\ell}}{T}- \dv{\sigma_{\nu_{\ell'}}}{T} = 
    \frac{\mathrm{G}_\mathrm{F}^2 M_\mathrm{A}}{4\pi} \left( 1-\frac{T}{E_\nu} -\frac{M_\mathrm{A} T}{2E_\nu^2} \right) &\qty[ \frac{2\alpha}{\pi} (\delta^{\nu_\ell}-\delta^{\nu_{\ell'}} ) \mathrm{F}_\mathrm{W} \left(Q^2\right) \mathrm{F}_\mathrm{ch}(Q^2) ]+ \mathrm{O}(\mathrm{G}_\mathrm{F}^2\alpha^2).
\end{split}
\end{equation}
Hadronic uncertainties, being proportional to $\delta^{\text{QCD}}$, cancel identically, whereas leptonic loops contribute differently with their difference being proportional to $(\delta^{\nu_\ell}-\delta^{\nu_{\ell'}} )$, given explicitly by
\begin{equation}\begin{split}
    \delta^{\nu_\ell}-\delta^{\nu_{\ell'}} &= 2\qty[ \Pi(Q^2, m_\ell)-\Pi(Q^2, m_\ell')]~ +\mathrm{O}( \alpha ),
\end{split}\end{equation}
 where we dropped the renormalization-scale dependence of the vacuum polarizations since the difference is renormalization scale and scheme independent.

\subsection{Next-to-leading order prediction \label{NLO-diff}}

Because the flavor asymmetry is $\mathrm{O}(\mathrm{G}_\mathrm{F}^2\alpha)$ at leading order itself, we can compute it up to next order, i.e. up to $\mathrm{O}(\mathrm{G}_\mathrm{F}^2\alpha^2)$. For this purpose, the ratio of the difference of flavor-dependent cross sections to a particular flavor cross section (e.g. $\sigma_{\nu_\mu}$) can be calculated to a higher level of accuracy than the difference alone.

According to Furry's theorem, contributions with two photons attached to the nucleus vanish. All QED vacuum polarization contributions to the photon line in~\cref{fig:closed_loop_ET} are captured replacing the overall coupling constant by its value in the Thomson limit $\alpha \to \alpha_0$. The only remaining flavor-dependent contributions at order $\mathrm{O}(\mathrm{G}_\mathrm{F}^2\alpha^2)$ arise from QED corrections to closed lepton loops in~\cref{fig:closed_loop_ET}. We include this correction taking analytical expressions from~\cite{Djouadi:1987gn,Djouadi:1987di,Kniehl:1989yc,Fanchiotti:1992tu,Tomalak:2019ibg}.

The resulting flavor asymmetry between $\nu_{\ell}$ and $\nu_{\ell'}$ cross sections is given by 
\begin{equation}
    \frac{\mathrm{d} \sigma_{\nu_\ell} - \mathrm{d} \sigma_{\nu_{\ell'}}}{\mathrm{d} \sigma_{\nu_\ell}} = r - r^2 + \mathrm{O} \left(\alpha^3 \right), \quad r = 4\frac{\alpha_0}{\pi}  \frac{ \left( \mathrm{\Pi} \left( Q^2,m_\ell \right)  - \mathrm{\Pi} \left( Q^2,m_{\ell'} \right)\right)  \mathrm{F}_\mathrm{ch} \left(Q^2\right)}{\mathrm{F}_\mathrm{W} \left(Q^2\right)  + \frac{\alpha_0}{\pi} \qty[\delta^{\nu_\ell} + \delta^\text{QCD}]   \mathrm{F}_\mathrm{ch}(Q^2)}. \label{eq:flavor_difference}
\end{equation}

\subsection{Uncertainties for a monoenergetic source \label{error-buget-asym}}

In this section, we focus on the flavor asymmetry defined in~\cref{eq:flavor_difference} which can be reliably computed to next-to-leading order. Moreover, the relative error for the flavor asymmetry is actually \emph{lower} than the relative uncertainty of the total cross section as can be seen by comparing~\cref{tab:errors_Argon} vs~\cref{tab:errors_Argon_difference,tab:errors_Argon_difference_tau}.

The reason for the lessened uncertainty can be understood as follows. Most of the nuclear uncertainties cancel in the ratio drastically decreasing the relative uncertainty at higher energies of $\mathrm{O}(50~\text{MeV})$. The hadronic contribution cancels in the numerator, and only enters at next-to-leading order because of its' entering the cross section in the denominator. The same holds true for the quark couplings error.

\begin{table}[hb]
\centering
\footnotesize
\centering 
\begin{tabular}{|c|c||c|c|c|c|c|c|}   
\hline          
$E_\nu,~\mathrm{MeV}$  &  Asymmetry &  Nuclear & Nucleon  & Hadronic & Quark  & Perturbative & Total 
\\
\hline
50  & 0.93 & 0.82   &  0.002  & 0.28 & 0.07 & 0.001  & 0.87
\\
30  & 1.14 & 0.21  & $9\times 10^{-4}$ & 0.28 & 0.07 & 0.001  & 0.35
\\
10 & 1.67 & 0.02 & $2\times 10^{-4}$  & 0.28 & 0.07 & 0.001 & 0.29
\\
\hline
\end{tabular}
\caption{\cevns flavor asymmetry, $(\sigma_{\nu_e}- \sigma_{\nu_\mu})/\sigma_{\nu_\mu}$ in $\%$, on $~^{40} \mathrm{Ar}$ target and contributions to the relative error (in $\%$) for an incident neutrino energy $E_\nu$.
\label{tab:errors_Argon_difference} } 
\end{table}

\begin{table}[hb]
\centering
\footnotesize
\centering 
\begin{tabular}{|c|c||c|c|c|c|c|c|}   
\hline          
$E_\nu,~\mathrm{MeV}$  &  Asymmetry &  Nuclear & Nucleon  & Hadronic & Quark  & Perturbative & Total 
\\
\hline
50  & 1.47 & 0.43   &  0.007  & 0.28 & 0.07 & 0.002  & 0.52
\\
30  & 1.47 & 0.17  & 0.003 & 0.28 & 0.07 & 0.002  & 0.34
\\
10 & 1.47 & 0.02 & $3\times 10^{-4}$  & 0.28 & 0.07 & 0.002 & 0.29
\\
\hline
\end{tabular}
\caption{\cevns flavor asymmetry, $(\sigma_{\nu_\tau}- \sigma_{\nu_\mu})/\sigma_{\nu_\mu}$ in $\%$, on $~^{40} \mathrm{Ar}$ target and contributions to the relative error (in $\%$) for an incident neutrino energy $E_\nu$.\label{tab:errors_Argon_difference_tau} } 
\end{table}

\begin{figure}
    \centering
    \includegraphics[height=0.226\textwidth]{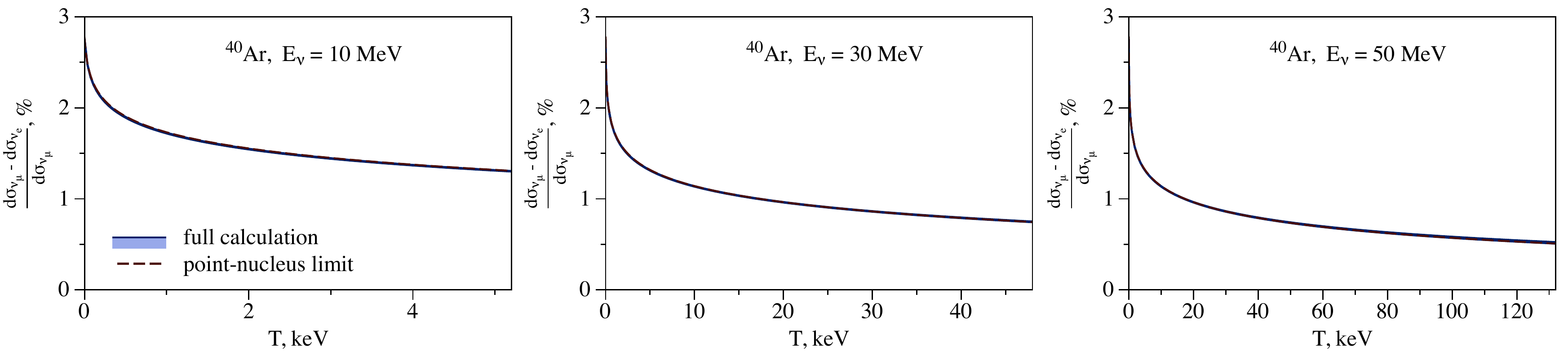}
    \caption{Same as~\cref{fig:CEvNS_Argon} but for the ratio $\left(\dd\sigma_{\nu_\mu}-\dd\sigma_{\nu_e}\right)/\dd\sigma_{\nu_\mu}$ compared to the point-nucleus limit of~\cref{sec:point_nucleus}. The exact and approximate calculations lay on top of each other. As discussed in~\cref{NLO-diff} by taking the difference of flavor-dependent cross sections the leading-order QCD corrections, being flavor independent, cancel in the numerator, while many nuclear systematic uncertainties cancel in the ratio. The result is a much smaller relative error compared to the absolute cross section as demonstrated in~\cref{tab:errors_Argon_difference,tab:errors_Argon_difference_tau}.\label{fig:CEvNS_Argon_differences}}       
\end{figure} 
\begin{figure}
    \centering
    \includegraphics[height=0.226\textwidth]{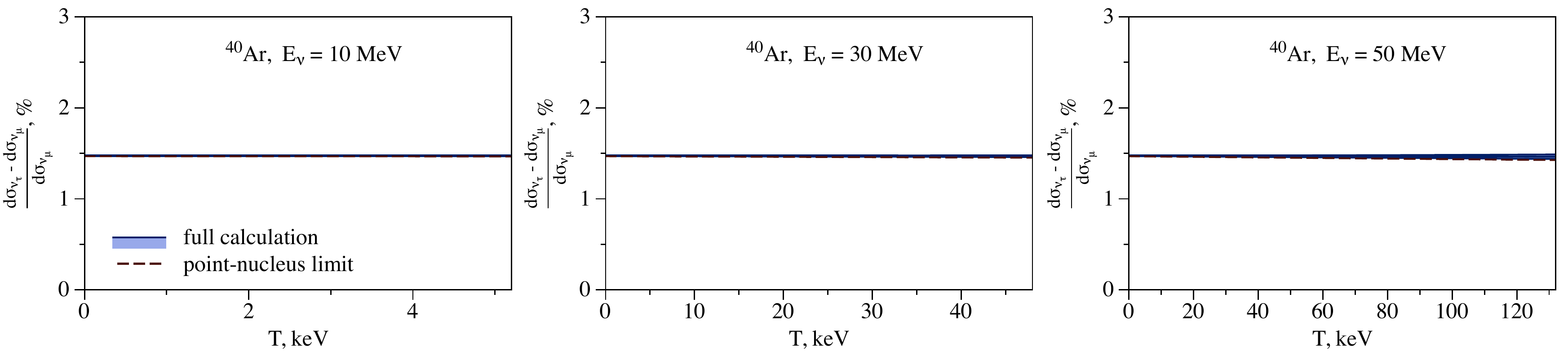}
    \caption{Same as~\cref{fig:CEvNS_Argon_differences} but for the ratio $\left(\dd\sigma_{\nu_\tau}-\dd\sigma_{\nu_\mu}\right)/\dd\sigma_{\nu_\mu}$. The $Q^2$ dependence in kinematics of $\pi$DAR arises mainly from closed electron loops and has to be included for flavor differences with electron flavor involved. \label{fig:CEvNS_Argon_differences_tau}}
\end{figure}

\subsection{Point-nucleus limit}
\label{sec:point_nucleus}
For the reader's convenience, we provide explicit formula in the idealized limit of a point-like nucleus $R_p, R_n\rightarrow 0$. Note that this limit is equivalent to $Q^2\to0$ only at tree level due to the $Q^2$ dependence of the vacuum polarization.
In this limit, the nuclear form factors assume their point-like values $\mathrm{F}_\text{ch} \rightarrow Z$ and $\mathrm{F}_\text{W} \rightarrow Q_\text{W}$ and the next terms in the $Q^2$ expansion can be neglected, i.e. $R_p E_\nu,~R_n E_\nu \ll 1$, such that
\begin{equation}
    \lim_{R_p,R_n\rightarrow 0}  \dv{\sigma_{\nu_\ell}}{T}- \dv{\sigma_{\nu_{\ell'}}}{T} =
     \frac{Z \alpha_0}{\pi} \qty[ \Pi(Q^2,m_\ell)-\Pi(Q^2,m_{\ell'})] \frac{\mathrm{G}_\mathrm{F}^2 M_\mathrm{A}}{\pi} \left( 1-\frac{T}{E_\nu} -\frac{M_\mathrm{A} T}{2E_\nu^2} \right)  
     Q_\text{W}, \label{eq:point_nucleus}
\end{equation}
where $ Q^2 = 2 M_A T$. We present the point-nucleus limit for relative flavor differences
\begin{equation}
   \lim_{R_p,R_n\rightarrow 0}  \frac{\mathrm{d} \sigma_{\nu_\ell} - \mathrm{d} \sigma_{\nu_{\ell'}}}{\mathrm{d} \sigma_{\nu_\ell}} =  4\frac{\alpha_0}{\pi}  \frac{Z}{Q_\mathrm{W}} \qty[ \mathrm{\Pi} \left( Q^2,m_\ell \right)  - \mathrm{\Pi} \left( Q^2,m_{\ell'} \right)], \label{eq:flavor_difference_point}
\end{equation} 
in~\cref{fig:CEvNS_Argon_differences,fig:CEvNS_Argon_differences_tau}. Our precise calculation is well approximated by~\cref{eq:flavor_difference}. In the kinematics of experiments with $\pi\mathrm{DAR}$ beams, the vacuum polarization function can be taken at $Q^2 = 0$ for tau neutrinos and, to a reasonably good approximation, for muon neutrinos. Within such an approximation, the flavor asymmetry reads as
\ber
   \lim_{R_p,R_n\rightarrow 0}  \frac{\mathrm{d} \sigma_{\nu_\tau} - \mathrm{d} \sigma_{\nu_{\mu}}}{\mathrm{d} \sigma_{\nu_\mu}} =  \frac{4}{3}\frac{\alpha_0}{\pi}  \frac{Z}{Q_\mathrm{W}} \ln \frac{m^2_\tau}{m^2_\mu}, \label{eq:flavor_difference2}
\eer
in agreement with the effective Weinberg angle convention as it is described in~\cref{prescription-app}.

%

\section{Applications\label{applications}}
%

The motivation for the precise calculation performed here is the rapid progress in technology for \cevns detector. 
In the three years since its discovery, the field of \cevns research has blown up and we optimistically await improved detector technologies and larger exposures that will drive down systematical and statistical uncertainties. 
With this outlook in mind, we briefly outline some potentially interesting applications of the work presented here that we hope will motivate and inform future experiments.

\subsection{Precision electroweak observables at low energy \label{precision-EW}}

To test the Standard Model of particle physics at low energies, our precise predictions of cross sections with quantified uncertainties~\cite{Tomalak:2019ibg,Hill:2019xqk} can be directly compared to experimental measurements. Any significant deviations between the theory and experiment will indicate the presence of new physics and can be expressed as some conventional definition for the low-energy property, traditionally referred to the Weinberg angle or the neutrino charge radius. More generally, what can be probed is the weak form factor of a nucleus and the neutrino’s electromagnetic form factor. Recent work has considered \cevns as a percent-level probe of physics both within~\cite{Scholberg:2005qs,Canas:2018rng,Khan:2019cvi,Huang:2019ene,Miranda:2019skf,Cadeddu:2019qmv,Baxter:2019mcx,Papoulias:2019xaw,Fernandez-Moroni:2020yyl,Bernabeu:2002nw,Bernabeu:2002pd,Papavassiliou:2005cs,Kosmas:2017tsq,Cadeddu:2018dux,Cadeddu:2019eta} and beyond the Standard Model (BSM) ~\cite{Aguilar-Arevalo:2019jlr,Amanik:2004vm,Barranco:2005yy,Scholberg:2005qs,Barranco:2007tz,Barranco:2008rc,Barranco:2009px,Formaggio:2011jt,Lindner:2016wff,Akimov:2017ade,Coloma:2017ncl,Kosmas:2017tsq,Liao:2017uzy,AristizabalSierra:2017joc,Farzan:2017xzy,Boehm:2018sux,Billard:2018jnl,Altmannshofer:2018xyo,AristizabalSierra:2018eqm,Brdar:2018qqj,Abdullah:2018ykz,Farzan:2018gtr,Denton:2018xmq,Gonzalez-Garcia:2018dep,Esteban:2018ppq,AristizabalSierra:2019zmy,Miranda:2019skf,Alikhanov:2019drg,Bell:2019egg,Bischer:2019ttk,AristizabalSierra:2019ufd,Dev:2019anc,Khan:2019cvi,Cadeddu:2019eta,Giunti:2019xpr,Akimov:2019rhz,Han:2019zkz,Aguilar-Arevalo:2019zme,Baxter:2019mcx,Papoulias:2019xaw,Coloma:2019mbs,Babu:2019mfe,Canas:2019fjw,Dent:2019ueq,Dutta:2019eml,Miranda:2019wdy,Esteban:2020opq,Denton:2020hop,Dutta:2020che,Chang:2020jwl,Flores:2020lji,Akimov:2020pdx,Abdullah:2020iiv,Li:2020lba,Amaral:2020tga,Abdallah:2020biq,Sadhukhan:2020etu,Suliga:2020jfa,Sinev:2020bux,Skiba:2020msb,Dutta:2020enk,Denton:2020uda,Ding:2020uxu,Cadeddu:2020nbr,Miranda:2020syh}.
Our precise theoretical treatment will allow us to consistently account for theoretical errors putting constraints on standard and non-standard neutrino couplings either on more specific parameters or concrete BSM scenarios.

We would like to stress that any analysis of low-energy properties at percent level or better has to be supplemented with a complete treatment of electroweak and QCD virtual corrections, at least at one-loop level. In particular, neutrino-electron scattering at energies of accelerator neutrinos and coherent elastic neutrino-nucleus scattering require a complete account for the kinematic dependence of radiative corrections which is often overlooked in the literature.

\begin{figure}[h]
          \centering
          \includegraphics[height=0.69\textwidth]{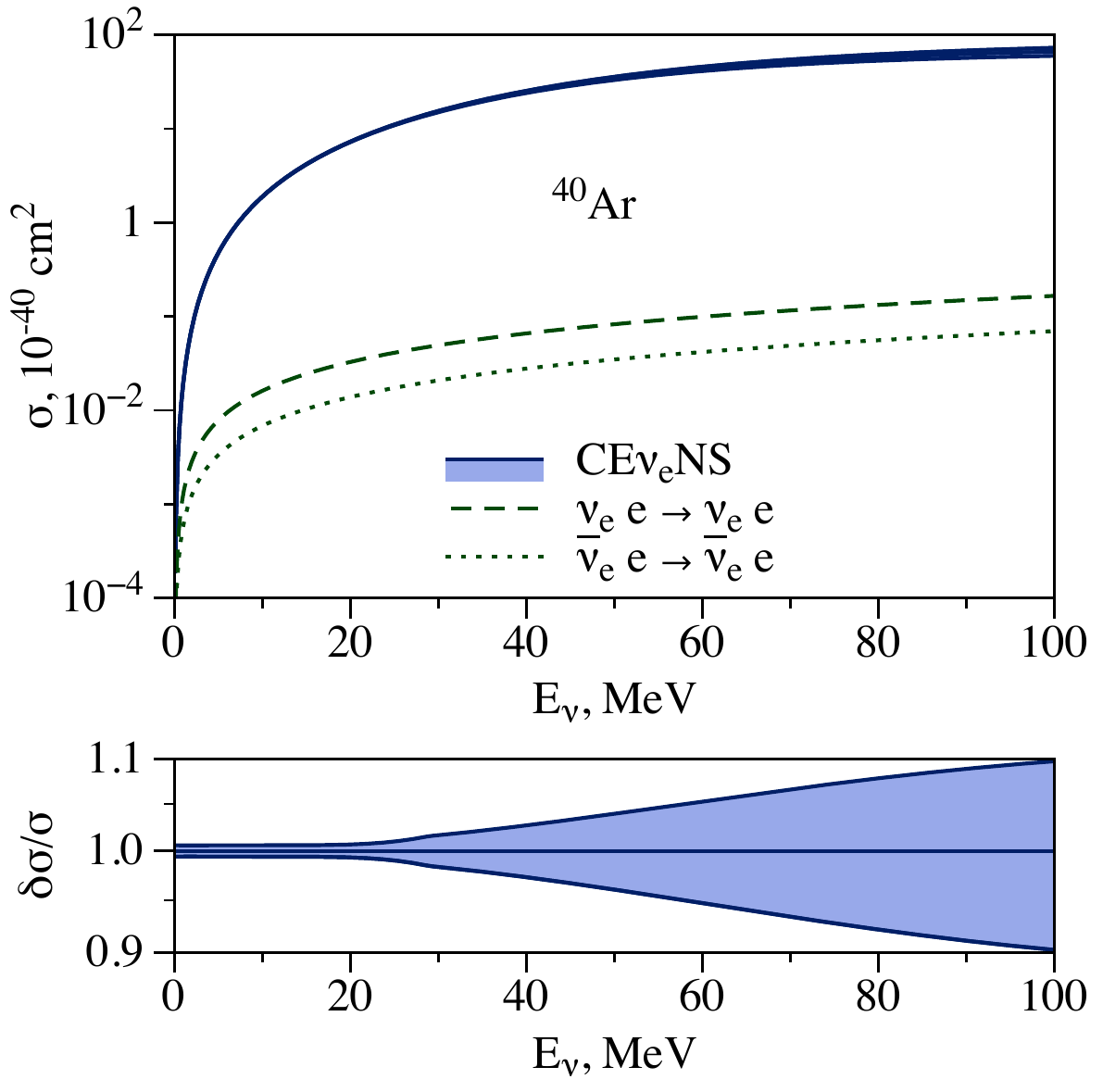}              
          \caption{On the upper plot, CE$\nu_e$NS cross section on $^{40}\mathrm{Ar}$ (blue band) is compared to $\nu_e e$ and $\bar{\nu}_e e$ cross sections of~\cite{Tomalak:2019ibg} multiplied by the factor $Z=18$. On the lower plot, the relative error of the total CE$\nu_e$NS cross section is presented. \label{fig:Ar_nue_xsections}}
\end{figure}

Working with correct (i.e.\ physical and propagating) degrees of freedom at low energies is the other crucial physical feature that is typically not taken into account. 
Quarks are not valid degrees of freedom at and below the QCD scale $\Lambda_\mathrm{QCD}$. One has to switch to the hadronic description which introduces additional uncertainty in any neutral-current neutrino-induced process due to the pure knowledge of the charge-isospin current-current correlation function $\hat{\Pi}_{3\gamma}$. 
The knowledge of $\hat{\Pi}_{3\gamma}$ at low energies can be improved by a factor 5-10 with precise lattice QCD studies~\cite{Burger:2015lqa,Ce:2018ziv,Ce:2019imp}. We would like to stress that hadronic physics introduces an error at a few permille level that is universal to \emph{all} neutrino-induced neutral-current scattering.
Going to low energies where nuclear physics is under control, these universal uncertainties dominate the error budget for both \cevns and neutrino-electron scattering. 
Therefore, we conclude that \cevns is actually an ideal tool for precision tests of electroweak physics due to the relative ease with which high statistics samples can be obtained (since the \cevns cross section is much larger than the $\nu e$ cross section). In~\cref{fig:Ar_nue_xsections}, to illustrate how much larger the \cevns cross section is, we compare the CE$\nu_e$NS cross section on $^{40}\mathrm{Ar}$ with neutrino-electron cross sections of~\cite{Tomalak:2019ibg} multiplied by the factor $Z=18$.
The main drawbacks for performing precision physics with \cevns are experimental uncertainties such as quenching factor and background rejection and the uncertainty on the neutrino flux.
While we expect future experiments to mitigate the former, the latter is an issue shared by almost all neutrino sources. 

\subsection{Prompt to delayed ratios with decay-at-rest sources}
One particularly interesting application of our results involves the prompt to delayed ratio that is often considered at $\pi$DAR experiments. At COHERENT, for example, the finite time required for a beam spill limits the efficiency of a timing cut to being $\mathrm{O}(10\%)$~\cite{COHERENT-FLUX}, however this is a well-studied effect, and it is reasonable to expect the collaboration to understand the precision of this efficiency at the percent level. Such a timing cut generates two samples of events: prompt and delayed, the former being composed of $\mathrm{O}(90\%)$ $\nu_\mu$ and $\mathrm{O}(10\%)$ $\bar{\nu}_\mu$, $\nu_e$ and vice-versa for the latter. 
 As we have demonstrated above, the $\nu_e$ and $\nu_\mu$ \cevns cross sections differ at the level of a few percent, and the two event samples would therefore be expected to yield event rates that also differ from the naive 1:2 event ratio at the percent level. 
 Since ratios of prompt to delayed fluxes are a standard ``handle'' for $\pi$DAR data, understanding this effect in detail for a given experiment's timing efficiency is an important issue to address. 
 
This is particularly important to BSM scenarios such as nonstandard interactions (NSIs)~\cite{Denton:2018xmq,Coloma:2019mbs,Dutta:2020enk, Dutta:2019eml, Baxter:2019mcx, Abdullah:2020iiv, Denton:2020hop} and sterile neutrino oscillations~\cite{Blanco:2019vyp, Miranda:2019skf}. As described above, timing cuts allow one to probe the flavor structure of a $\pi$DAR source, and flavor-dependent NSIs are easily capable of explaining any asymmetry between the prompt and delayed samples. If similar studies are to be conducted with percent level precision at future experiments then proper accounting for the SM flavor asymmetry is once again mandatory.
 
 Finally, in addition to flavor-dependent corrections, the simple observation that the cross section is shifted relative to its SM predicted value is important for experiments with $\pi$DAR, nuclear reactor neutrino sources, and isotope decay-at-rest sources (IsoDAR)~\cite{Bungau:2012ys}.
 We will discuss the latter in the next section.
 A precise knowledge of the \cevns cross section could be coupled with an IsoDAR or $\pi$DAR source to provide high-precision calibrations of quenching factors~\cite{Foxe:2012ega,Rich:2017lzd,Bonet:2020awv,Lewis:2021cjv} which could then be used in applications with \cevns detectors at nuclear reactor sources where flux uncertainties are more severe, as we will see shortly.

\subsection{Isotope decay at rest \label{ISODAR} } 

As alluded to at the end of~\cref{error-budget}, the theoretical uncertainty of the \cevns cross section is incredibly small at low recoil energies being dominated by hadronic current-current correlators and uncertainty on quark-level couplings in the effective field theory description used here. 
Moreover, \cevns is easily the largest neutrino cross section available at low energies, surpassing both inverse beta decay and neutrino-electron scattering, and, as we have emphasized in this work, is even sensitive to flavor differences at the percent level. 
Therefore, were a high-intensity source of low-energy ($\sim 10$ MeV) neutrinos to exist, \cevns would represent a powerful tool with which to conduct precision experiments. 
One would have both large statistics and high-precision predictions.

Such a low-energy neutrino source has been recently proposed as part of the DAE$\delta$ALUS facility~\cite{Alonso:2010fs} being termed isotope decay at rest (IsoDAR)~\cite{Bungau:2012ys}. The basic premise is to irradiate a beryllium target, liberating neutrons, which are subsequently captured in a surrounding isotopically ($99.9\%$) pure $^7$Li sleeve. Neutron capture in the absorber leads to a large population of radioactive $^8$Li which then beta decays yielding a low-energy $\bar{\nu}_e$. The beta decay of $^8$Li has a relatively well-understood decay spectrum and an ongoing program of Monte Carlo studies are underway~\cite{Zhao:2015bba} to have an accurate prediction for the $\bar\nu_e$ flux. 

The produced $\bar{\nu}_e$ flux can then be detected via either elastic scattering or inverse beta decay (IBD)\footnote{For concreteness, we restrict inverse beta decay to the capture of electron antineutrinos on free protons.} $\bar{\nu}_e p \rightarrow n e^+$. A high-intensity IsoDAR source has many of the same advantages of a nuclear reactor source, without the complicated problem of uncertain and time-dependent fuel composition. The energy is sufficiently high that IBD thresholds are easily overcome. Because the energy of the signal $e^+$ from IBD is highly correlated with antineutrino energy, many IsoDAR studies focus on its ability to yield a large flux of neutrinos with measurable energy.

Other complementary detector strategies for IsoDAR have also been proposed including \cevns and elastic neutrino-electron scattering. The latter of these two channels has a very small cross section, but is often touted as a perfect setting in which to perform precision tests of the Standard Model at low energies~\cite{Adelmann:2013isa,Conrad:2013sqa}.
While electrons are naively a perfectly ``clean'' target, as we discuss in~\cref{precision-EW} at $\mathrm{O}(\mathrm{G}_\mathrm{F}^2\alpha)$, the same charge-isospin hadronic correlator, introduced in~\cref{eq:radiative_correction-3}, enters the radiative corrections and dominates the theoretical uncertainty of the cross section. In~\cite{Adelmann:2013isa}, both a \cevns detector and a $\nu e$ scattering detector were proposed on the merits of observing nuclear coherence and conducting precision electroweak measurements respectively. At IsoDAR energies, however, we find that the SM prediction for the \cevns cross section is competitive with the relative precision of the $\nu e$ scattering cross section, but would allow for a much larger statistical sample. 
We therefore conclude that a \cevns detector could perform both tasks at an IsoDAR source.

\subsection{Standard oscillations} 
As flavor corrections lead to slightly different \cevns cross sections for each neutrino flavor, it raises the following question: ``Can oscillation physics be studied with CE$\nu$NS''?
The main advantage here compared to other neutrino oscillation experiment is that \cevns has no particle production threshold.
Therefore, one may study tau neutrino physics well below the tau production threshold of about $3.5$~GeV.
For example, with a $\pi$DAR source the maximum of tau appearance would require about 15~km baseline.
In fact, exploring \cevns to study tau neutrinos would be the only way currently known to study $\nu_\tau (\bar{\nu}_\tau)$ oscillations with (statistical) flavor identification below the tau threshold.

Another advantage would be related to the size of the detector necessary to study neutrino oscillations.
First, the large \cevns cross section would allow for smaller detectors to see a relatively large number of events.
Second, the shorter baseline to observe oscillations, compared to e.g. 810~km and 295~km of the current oscillation experiments NOvA~\cite{Ayres:2007tu} and T2K~\cite{Abe:2011ks}, would also help getting more events.
The obvious disadvantage lies in the fact that the flavor-dependent effects in \cevns are at a few percent level, and thus the changes in event counts induced by oscillations are typically small.

To have an idea of what could be done, in principle, with a \cevns detector for standard oscillation physics, we have performed the following simple exercise. 
To estimate the statistical precision one may achieve in such a setup, we have (very optimistically) assumed that a total liquid argon detector and $\pi$DAR source such that $(\text{total number of pions produced})\times(\text{detector fiducial mass}) = 20 N_\mathrm{A}\,\text{kton}$, where $N_\mathrm{A}$ is the Avogadro number.
If the detector is deployed 15~km from the source, 
the prompt monochromatic line would lead to about 10,000 \cevns interactions while the delayed continuous spectrum would lead to 30,000 events. 
This translates into 1\% and 0.6\% statistical uncertainties.
The relative difference in the number of events between unoscillated and oscillated neutrino fluxes would be 1.3\% and 0.6\% for the prompt and delayed neutrinos, assuming the best fit oscillation parameters of~\cite{Esteban:2020cvm} and using the flavor-dependent \cevns cross section of~\cref{eq:flavor_difference_point}.

To obtain some understanding of these numbers, we first note that the prompt flux of $\nu_\mu$ almost entirely oscillates to $\nu_\tau$ at 15~km.
While a similar oscillation effect happens to the delayed $\bar\nu_\mu$ component, $\nu_e$ disappearance is driven by the smallest mixing angle $\sin^22\theta_{13}\simeq 0.09$. 
Therefore, the oscillation effect on the delayed component is about half of that in the prompt events, modulus the energy dependence of the delayed flux and $Q^2$ dependence on asymmetries involving the electron flavor.
Also, contributions to the asymmetry by $\nu_\mu\to\nu_\tau$ and $\nu_\mu\to\nu_e$ oscillations have the opposite sign.

From these numbers, one could, within the proposed optimistic scenario, measure tau appearance below the tau threshold at the couple of $\sigma$ level.
Note also that, while a near detector would certainly be helpful, the ratio between prompt and delayed spectrum could mitigate the uncertainties on the initial neutrino flux.

Doing precision oscillation physics with such a setup would be quite a challenge: percent level sensitivity to e.g. $\sin^22\theta_{23}$ would require at least a statistical uncertainty of {\rm a few}~\%, which seems a bit too optimistic (that is, unrealistic) for next, or even next-to-next generation experiments.
Nevertheless, studying tau neutrinos with this setup could provide nontrivial information on the unitarity of the PMNS matrix~\cite{AguilarSaavedra:2000vr,Farzan:2002ct,He:2013rba,Ellis:2020ehi,Ellis:2020hus}.
As a final remark, future dark matter direct detection experiments like DARWIN~\cite{Aalbers:2020gsn} will be able to observe \cevns from the $^8$B solar neutrino flux.
The expected number of \cevns events from this flux above the threshold is about 90 events per ton-year~\cite{reichard_2020} which would translate into 36,000 events for a 40 ton dual-phase xenon detector running for 10 years. 
The number of events could be increased if the nuclear recoil detection threshold is lowered, possibly allowing for oscillation studies at DARWIN with CE$\nu$NS.
To understand if a realistic experimental setup could leverage the flavor dependence of the \cevns cross section to perform oscillation physics studies is beyond the scope of this paper and may be pursued in the future.

\subsection{Oscillations to sterile neutrinos}

The LSND~\cite{Aguilar:2001ty} and MiniBooNE~\cite{Aguilar-Arevalo:2018gpe} anomalies present an outstanding conundrum in the field of neutrino physics. 
While these anomalies are consistent with the existence of light sterile neutrinos at the eV, such an interpretation presents severe tension with numerous experiments~\cite{Dentler:2017tkw, Dentler:2018sju, Diaz:2019fwt, Boser:2019rta} and with standard $\Lambda$CDM cosmology~\cite{Abazajian:2013oma,Chu:2015ipa,Bridle:2016isd,Aghanim:2018eyx}. The sterile neutrino searches in most oscillation experiments primarily rely on charged-current interactions using either an appearance signal, interpreted as an active flavor oscillating to another active flavor, or a disappearance signal, interpreted as an active flavor oscillating to any other neutrino flavor (active or sterile). 

Oscillation probabilities are simplest for monoenergetic sources and $\pi$DAR provides a high-intensity monoenergetic source of $\nu_\mu$, thus being a natural candidate for carrying out high statistics searches for $\nu_\mu$ disappearance. 
Importantly, however, the daughter neutrino from $\pi$DAR is well below the charged-current threshold required to produce muons, and so one must rely on neutral-current processes. In this context, \cevns has been recognized as being advantageous due to its relatively large cross section~\cite{Formaggio:2011jt, Anderson:2012pn, Kosmas:2017zbh, Miranda:2020syh}. Working at lower energies allows for shorter baselines with equivalent $L/E_\nu$ and consequently higher fluxes as compared to e.g.\ the SBN program.

Regarding the experimental landscape, currently a 10-ton liquid argon experiment, Coherent CAPTAIN-Mills (CCM) at Lujan center at LANL, is operational and plans to study active-to-sterile neutrino oscillations using \cevns~\cite{CCM} at multiple baselines from the $\pi$DAR source. 
Future measurements with ton and multi-ton scale \cevns detector at the Spallation Neutron Source at ORNL~\cite{Akimov:2019xdj} and at the European Spallation Source~\cite{Baxter:2019mcx} are at planning stage. 
A key assumption in all of these proposals, however, is that \cevns is flavor independent. 
As we have shown in this work, \cevns has a \emph{precisely calculable} dependence on neutrino flavor that enters at a few percent level and is especially pronounced at low nuclear recoil energies. 
If CCM or similar experiments are to search for disappearance probabilities on the order of a few percent, these SM radiative corrections become obligatory for a proper analysis. Even with a near detector, flavor-dependent corrections to the \cevns cross section are required to properly interpret precise experimental results.

\subsection{Nuclear facility monitoring \label{monitoring}}

Recent work has shown that neutrino detectors, both using IBD (i.e. $\bar{\nu}_e$ capture on free protons) and \cevns on nuclei, can serve as novel state of the art instruments for monitoring nuclear facilities (see~\cite{Bernstein:2019hix} for a review). This includes both civilian nuclear reactors, where precise measurements of the $\bar{\nu}_e$ flux can provide information about spent fuel~\cite{Brdar:2016swo,Jaffke:2016xdt}, and the global excess of fissile nuclear materials~\cite{Christensen:2013eza,Cogswell:2016aog}. These ideas have already motivated the construction of one such neutrino-based nuclear reactor monitoring system~\cite{Alfonzo:2018zpz}.

In both cases, \cevns occupies a unique and complementary position relative to IBD because of its thresholdless nature~\cite{Cogswell:2016aog,Bowen:2020unj}. There are two ways in which \cevns differentiates itself from IBD. First, the cross section is much larger. This observation holds true even after normalizing per unit of detector mass~\cite{Bowen:2020unj} provided the thresholds are low enough (sub 100 eV). Perhaps more important, however, is the fact that certain neutrino flux components appear \emph{only} below the IBD threshold, $E_\nu \lesssim 1.8$ MeV, and so \cevns is \emph{the only} tool available for certain applications. A notable example is the monitoring plutonium blankets~\cite{Cogswell:2016aog} as a marker of compliance with the Plutonium Management and Disposition Agreement.\footnote{Agreement between the Government of the United States of America and the Government of the Russian Federation concerning the management and disposition of plutonium designated as no longer required for defense purposes and related cooperation,\href{http://fissilematerials.org/library/2010/04/2000_plutonium_management_and_.html}{http://fissilematerials.org/library/2010/04/2000\_plutonium\_management\_and\_.html}.} Interestingly, it is in this low-energy, low-nuclear recoil limit where radiative corrections are most sizeable as can be clearly seen by comparing the three panels of~\cref{fig:CEvNS_Argon}. Moreover, (anti)neutrinos emitted in nuclear reactions are electron flavor eigenstates for whom radiative corrections are the largest (relative to muon and tau neutrino flavors).

Neutrino energies substantially below the IBD threshold may be accessible in the future by employing silicon-based Skipper-CCDs, which have recoil thresholds in the 10-20 eV range. The most abundant isotope of silicon, $^{28}$Si, composes over 90\% of a naturally occurring silicon source and is, fortuitously, a spin-0 nucleus. The other two stable isotopes, $^{29}$Si and $^{30}$Si, are spin-1/2 and spin-0 and have natural abundances of 5\% and 3\% respectively. Although beyond the scope of this work, the effective field theory treatment can be extended to spin-1/2 nuclei. 

The level of precision achieved here paves the way for high-precision neutrino detection of nuclear facilities. Having provided a permille level error budget for the \cevns cross section, we essentially guarantee that near-future neutrino probes of nuclear reactors will be limited by either statistics or experimental systematics. \cevns naturally complements IBD because of its thresholdless nature allowing detectors to measure components of the neutrino flux with $E_\nu \lesssim 1.8$ MeV~below IBD thresholds. Some of these low-energy components are essential for applications~\cite{Cogswell:2016aog}. To measure low-energy neutrinos necessarily requires a low nuclear recoil threshold, and it is interesting to note that the radiative corrections studied in this work are largest in the low-recoil limit. We hope that the ability to conduct future precision measurements of $\bar{\nu}_e$ spectra uninhibited by theoretical errors will serve as a useful tool in the monitoring of civilian and military nuclear facilities. 

\section{Conclusions and outlook \label{conclusions}}

We have provided a comprehensive treatment of \cevns cross sections on spin-0 nuclei appropriate for next-to-leading order accuracy of both the overall flavor-dependent cross section and flavor asymmetries. Our calculation accounts for all sources of theoretical errors including nuclear form factors, nucleon form factors, perturbative error in the determination of low-energy Wilson coefficients, and nonperturbative hadronic contributions to radiative corrections. Extension to higher spin nuclei is possible after a careful account for nuclear responses.

Surprisingly, the largest source of uncertainty at low energies comes from hadronic physics and perturbative error in the determination of Wilson coefficients. At larger energies, nuclear form factor uncertainties associated with the distribution of neutrons dominate. Practically speaking, this means that the \cevns cross section for low-energy neutrinos is a precision observable on par with neutrino-electron scattering whose dominant theoretical uncertainty is driven by the same hadronic charge-isospin correlator as the \cevns cross section. 

Our results pave the way for future high-precision \cevns experiments. While technical hurdles must still be overcome to begin probing percent-level effects and to reach enhanced low-recoil cross sections, once these have been achieved \cevns will be able to test SM prediction of neutrino interactions and to search for new physics that leads to percent level (or optimistically permille level) deviations from SM predictions. We think that a \cevns detector coupled with an IsoDAR source is especially promising. However, a $\pi$DAR source could provide similarly high levels of precision provided the nuclear uncertainties that enter at higher neutrino energies are reduced to a permille level. 

In summary, we have provided a state of the art precision calculation of the \cevns cross section and flavor asymmetries that we hope will enable and motivate future experimental progress.

\vfill 
\pagebreak

\appendix

\section{Field redefinition for the effective Lagrangian \label{field-redefinition} }

In this paper, we have emphasized the role of the photon in \cevns but our effective Lagrangian does not contain any explicit neutrino-photon coupling. As discussed in the main text, the absence of an explicit photon-neutrino coupling is a consequence of field redefinition which induces an explicit vector-like (weak-electromagnetic) current-current interaction between neutrinos and the nucleus. In this section, we offer a brief discussion of this field redefinition; for a detailed discussion see~\cite{Hill:2019xqk}.

Prior to the field redefinition, the neutral-current interaction of neutrinos with quarks and leptons relevant for scattering cross sections at $\mathrm{O}(\mathrm{G}_\mathrm{F}^2\alpha)$ is described by the effective four-fermion Lagrangian~\cite{Weinberg:1967tq,Hill:2019xqk} (neglecting fermion kinetic terms)
\begin{align}  \label{pre-redef}
    {\cal L}_{\rm eff} &\supset - \sum_{\ell,\ell^\prime}  \bar{\nu}_\ell \gamma^\mu \mathrm{P}_\mathrm{L} \nu_\ell
    \, \bar{\ell}^\prime \gamma_\mu (c_\mathrm{L}^{\nu_\ell \ell^\prime} \mathrm{P}_\mathrm{L}
    + c_\mathrm{R}^{\nu_\ell \ell^\prime} \mathrm{P}_\mathrm{R}) \ell^\prime
    - \sum_{\ell,q}  \bar{\nu}_\ell \gamma^\mu \mathrm{P}_\mathrm{L} \nu_\ell \,
    \bar{q} \gamma_\mu (c_\mathrm{L}^{q} \mathrm{P}_\mathrm{L} + c_\mathrm{R}^{q} \mathrm{P}_\mathrm{R}) q
    \nonumber\\
    & \quad\quad -\frac14 F_{\mu\nu}F^{\mu\nu}~+e\sum_{\ell}  Q_\ell \bar{\ell} \gamma_\mu \ell A^\mu + e\sum_q Q_q \bar{q}\gamma_\mu q A^\mu
    - \frac{1}{e} \sum_{\ell} c^{\nu_\ell \gamma} \partial_\mu F^{\mu\nu} \bar{\nu}_\ell \gamma_\nu \mathrm{P}_\mathrm{L} \nu_\ell.
\end{align}
The fact that neutrino-photon coupling only appears at dimension-6 is a manifestation of the anapole-only neutrino-photon vertex appropriate for a massless (or massive and Majorana) neutrino~\cite{Giunti:2014ixa}. 

For neutral-current scattering applications, it is convenient to redefine the photon field\footnote{We remind the reader that S-matrix elements are unaffected by field redefinitions. The same answers would be obtained if photon-mediated diagrams were included explicitly. } such that neutrino-photon interactions are removed at NLO. The appropriate field redefinition to achieve this feature is
\begin{equation}\label{redef}
  A_\mu \rightarrow A_\mu + \frac{1}{e} \sum_{\ell} c^{\nu_\ell \gamma} \bar{\nu}_\ell \gamma_\mu \mathrm{P}_\mathrm{L} \nu_\ell~.
\end{equation}
Such a field redefinition introduces additional four-fermion operators as
\begin{align}      \label{effective_Lagrangian_quark-app-old}
    {\cal L}_{\rm eff} &\supset - \sum_{\ell,\ell^\prime}  \bar{\nu}_\ell \gamma^\mu \mathrm{P}_\mathrm{L} \nu_\ell
    \, \bar{\ell}^\prime \gamma_\mu (c_\mathrm{L}^{\nu_\ell \ell^\prime} \mathrm{P}_\mathrm{L}
    + c_\mathrm{R}^{\nu_\ell \ell^\prime} \mathrm{P}_\mathrm{R}) \ell^\prime
    - \sum_{\ell,q}  \bar{\nu}_\ell \gamma^\mu \mathrm{P}_\mathrm{L} \nu_\ell \,
    \bar{q} \gamma_\mu (c_\mathrm{L}^{q} \mathrm{P}_\mathrm{L} + c_\mathrm{R}^{q} \mathrm{P}_\mathrm{R}) q 
    \nonumber\\
    & \quad\quad -\frac14 F_{\mu\nu}F^{\mu\nu} + e \sum_{\ell} Q_\ell \bar{\ell} \gamma_\mu \ell A^\mu + e \sum_q Q_q \bar{q}\gamma_\mu q A^\mu \\
     &\quad\quad\quad+ \qty[\sum_{\ell'} Q_{\ell'} \bar{\ell}' \gamma_\mu \ell'] \qty[\sum_{\ell} c^{\nu_\ell \gamma} \bar{\nu}_\ell \gamma_\mu \mathrm{P}_\mathrm{L} \nu_\ell]+ \qty[\sum_q Q_q \bar{q}\gamma_\mu q ] \qty[\sum_{\ell} c^{\nu_\ell \gamma} \bar{\nu}_\ell \gamma_\mu \mathrm{P}_\mathrm{L} \nu_\ell].\nonumber
\end{align}
The bottom line can then be absorbed into modified definitions of the left- and right-handed couplings, such that the effective Lagrangian is written as%
\begin{align}      \label{effective_Lagrangian_quark-app-new}
    {\cal L}_{\rm eff} &\supset - \sum_{\ell,\ell^\prime}  \bar{\nu}_\ell \gamma^\mu \mathrm{P}_\mathrm{L} \nu_\ell
    \, \bar{\ell}^\prime \gamma_\mu (c_\mathrm{L}^{\nu_\ell \ell^\prime} \mathrm{P}_\mathrm{L}
    + c_\mathrm{R}^{\nu_\ell \ell^\prime} \mathrm{P}_\mathrm{R}) \ell^\prime
    - \sum_{\ell,q}  \bar{\nu}_\ell \gamma^\mu \mathrm{P}_\mathrm{L} \nu_\ell \,
    \bar{q} \gamma_\mu (c_\mathrm{L}^{q} \mathrm{P}_\mathrm{L} + c_\mathrm{R}^{q} \mathrm{P}_\mathrm{R}) q 
    \nonumber\\
    & \quad\quad -\frac14 F_{\mu\nu}F^{\mu\nu} + e \sum_{\ell} Q_\ell \bar{\ell} \gamma_\mu \ell A^\mu + e \sum_q Q_q \bar{q}\gamma_\mu q A^\mu~, 
\end{align}
which matches~\cref{effective_Lagrangian_quarks} of the main text. This shows explicitly how high-energy photon-mediated diagrams can be re-shuffled into modified left- and right-handed couplings at low energies.

\section{Comparison with effective Weinberg angle convention \label{prescription-app}}

The literature surrounding \cevns discusses flavor-dependent radiative corrections, often phrased as a measurement of the neutrino charge radius. In this appendix, we compare our systematic treatment to the prescription commonly presented in the literature. 

The prescription in the literature is to take a tree-level \cevns cross section and to make the following replacement~\cite{Bernabeu:2002pd,Papavassiliou:2005cs,Cadeddu:2018dux}, 
\begin{equation}
\begin{split}
    \sin^2\theta_\mathrm{W} &\rightarrow \sin^2\theta_\mathrm{W}- \frac{\alpha}{4\pi}\qty[1-\tfrac23\ln\frac{m_\ell^2}{M_W^2}],
\end{split}
\end{equation}
for (anti)neutrino flavor $\nu_\ell(\bar{\nu}_\ell)$.

An easy point of comparison is our expression for the difference of the differential cross sections. In the prescription advocated in the \cevns literature~\cite{Bernabeu:2002pd,Papavassiliou:2005cs,Cadeddu:2018dux}, the difference would be given as
\begin{equation}
    \dv{\sigma_{\nu_\ell}}{T}- \dv{\sigma_{\nu_{\ell'}}}{T} \approx 
    \frac{\mathrm{G}_\mathrm{F}^2 M_\mathrm{A}}{3\pi} \frac{Z \alpha}{\pi} \left( 1-\frac{T}{E_\nu} -\frac{M_\mathrm{A} T}{2E_\nu^2} \right) N \ln \frac{m_{\ell'}^2}{m_\ell^2}~, 
\end{equation}
which is the same answer one would get starting from~\cref{eq:point_nucleus} in $m_e,m_\mu,m_\tau \rightarrow \infty$ limit (with $m_e/m_\mu$ and $m_\mu/m_\tau$ fixed), such that a small $Q^2$ expansion is justified. 

For realistic values of the lepton masses, this prescription is insufficient for electron (and quite often for muon) flavor and misses crucial $Q^2$ dependence leading to an overprediciton of the $\nu_e-\nu_\mu$ flavor asymmetry by a factor as large as six at $Q^2 \simeq 100~\text{MeV}^2$. A similar prescription, much closer to our conclusion, is presented in~\cite{Sehgal:1985iu,Degrassi:1989ip} where the Weinberg angle is replaced by the full $Q^2$-dependent form factor. A low-$Q^2$ expansion is only permissible for $Q^2 \ll m_\ell^2$. 

\section{Flux-averaged cross sections \label{flux-avg}}

In this paper, we have presented precise SM predictions for \cevns cross sections, but have intentionally refrained from discussing the experimentally relevant question of event rates from flux-averaged cross sections. A prediction for the event rate in a given interval of recoil energy $T\in [T_\text{min},T_\text{max}]$ requires the cross section to be folded against a flux prediction. To predict the event rate at the few-\% level, both the cross section \emph{and} the flux must be known to the same level of precision. 

Consider a flux of neutrinos sourced by a $\pi^+$ beam that is stopped in a target. Often these sources are termed pion decay at rest ($\pi$DAR) but they inevitably include contamination from pion decay in flight ($\pi$DIF) as well. The pions that do decay produce $\mu^+$, which subsequently decay to $\bar{\nu}_\mu, {\nu}_e, e^+$. Both the pion and muon decays also receive radiative corrections at the level of a few-\%. If one would like to predict the event rate at such an experiment at a percent level of precision then one requires a prediction for the $\pi$DIF component, $\Phi_{\nu_\ell}^{\pi\text{DIF}}$, the leading order $\pi$DAR flux (neglecting radiative corrections) for each neutrino flavor, $\Phi_{\nu_\ell}^{(0)}$, and the radiative corrections to this flux for each neutrino flavor, $\Phi_{\nu_\ell}^{(1)}$. Adding all of these together, one would find the total flux arriving at the detector accurate up to corrections of the permille level.
\begin{equation}
    \Phi_\nu = \Phi_{\nu_\ell}^{(0)} + \frac{\alpha}{\pi} \Phi_{\nu_\ell}^{(1)}+ \Phi_{\nu_\ell}^{\pi\text{DIF}} + \mathrm{O}(\text{permille})~.
\end{equation}
The $\pi$DIF flux was a few percent of the leading-order $\pi$DAR flux at LSND~\cite{Athanassopoulos:1996ds} and is expected to supply an $\mathrm{O}(0.5\%)$ contamination at the SNS~\cite{COHERENT-FLUX}. We therefore count this at the same order as the radiative corrections for practical purposes, but ignore the $\mu^-$ capture discussed in~\cite{COHERENT-FLUX} for brevity's sake. This flux then must be folded against the neutrino energy-dependent cross section calculated in this paper
\begin{equation}
    \dv{\sigma_{\nu_\ell}}{T}= \dv{\sigma^{(0)}}{T} + \frac{\alpha}{\pi} \dv{\sigma^{(1)}_{\nu_\ell}}{T},
\end{equation}
where $\dd\sigma^{(1)}_{\nu_\ell}/\dd T$ is the correction to the tree-level cross section $\dd\sigma^{(0)}/\dd T$. $\dd\sigma^{(1)}_{\nu_\ell}/\dd T$ can be taken from the main text of this paper.

For rate predictions, $\mathrm{R}_{\nu_\ell}$, at the percent level, one needs
\begin{equation}
    \dv{\mathrm{R}_{\nu_\ell}}{T} = \int \dd E_{\nu_\ell}\qty[ \dv{\sigma^{(0)}}{T} \Phi_{\nu_\ell}^{(0)} +\frac{\alpha}{\pi}\dv{\sigma^{(1)}_{\nu_\ell}}{T} \Phi_{\nu_\ell}^{(0)} +\frac{\alpha}{\pi}\dv{\sigma^{(0)}}{T} \Phi_{\nu_\ell}^{(1)} + \dv{\sigma^{(0)}}{T} \Phi_{\nu_\ell}^{\pi\text{DIF}} ]~,\label{flux-fold}
\end{equation}
with the integrand evaluated at fixed nuclear recoil energy. Since $Q^2=2M_\mathrm{A} T$, the complicated functional dependence on $Q^2$ (\emph{c.f.}~\cref{vacuum_polarization}) does not effect the integration over $E_{\nu_\ell}$. The explicitly $E_\nu$-dependent prefactor in~\cref{cevns-cross-section} must be included in both the LO and NLO expressions for the \cevns cross section.

Using the expressions in this paper, the monochromatic $\pi$DAR flux, and the well known $\mu$DAR flux (all without radiative corrections), the first two terms in~\cref{flux-fold} can be calculated. To calculate the last two terms in the square brackets, predictions for the corrections to the leading-order $\pi$DAR and $\mu$DAR fluxes are needed, as is a prediction for the $\pi$DIF component of the flux. We leave a calculation of the $\pi$DAR and $\mu$DAR radiative corrections to future work. The $\pi$DIF component (and $\mu^-$ capture component if it's contribution is appreciable) requires a dedicated study for each experiment.

\acknowledgments

We thank Sonia Bacca, Charlie Payne, Junjie Yang and their collaborators for providing us with tables of point-nucleon form factors. We also thank Ivan Martinez-Soler and Yuber Perez-Gonzalez for useful discussions. The work of O.T.\ is supported by the Visiting Scholars Award Program of the Universities Research Association. O.T.\ thanks Richard J.\ Hill for collaboration on~\cite{Tomalak:2019ibg,Hill:2019xqk} which has motivated the main part of this work. We thank Martin Hoferichter, Achim Schwenk and Javier Men\'{e}ndez for useful correspondence regarding the inputs from nuclear physics and suggestions to relate our definitions to the established literature, Raza Sufian and Keh-Fei Liu for correspondence regarding contributions of strange quark to neutrino cross sections, and Silas Beane for an inspiring question regarding the point-nucleus limit. R.P.'s visit to Fermilab was supported by funds from the Intensity Frontier Fellowship. R.P.\ and O.T.\ thank the Fermilab theory group for their hospitality and collaborative research environment. O. T. acknowledges the theory group of Institute for Nuclear Physics at Johannes Gutenberg-Universit\"at Mainz for warm hospitality and support.
R.P.\ and O.T.\ were supported by the U.S. Department of Energy, Office of Science, Office of High Energy Physics, under Award Number DE-SC0019095. VP acknowledges the support from US DOE under grant DE-SC0009824. This manuscript has been authored by Fermi Research Alliance, LLC under Contract No. DE-AC02-07CH11359 with the U.S. Department of Energy, Office of Science, Office of High Energy Physics. FeynCalc~\cite{Mertig:1990an,Shtabovenko:2016sxi}, LoopTools~\cite{Hahn:1998yk}, JaxoDraw~\cite{Binosi:2003yf}, Mathematica~\cite{Mathematica} and DataGraph were extremely useful in this work.

\paragraph{Note added.} Near the end of this work, we learned that the COHERENT collaboration has performed a calculation including SM radiative corrections internally. This calculation remains unpublished~\cite{COHERENT-INTERNAL}, and the only description of its contents we are aware of is on pages 10 \& 11 of~\cite{Scholz:2017ldm}. Comparison of our results is of great interest.

\bibliographystyle{jhep}
\bibliography{cevns}

\end{document}